\documentclass[manusacript,nonacm,natbib=false]{acmart}
\usepackage{hyperref}
\usepackage{subcaption}
\usepackage{array}
\usepackage{ragged2e}
\usepackage{makecell}
\newcolumntype{L}[1]{>{\RaggedRight\arraybackslash}p{#1}}
\usepackage{tcolorbox}
\usepackage{xurl}

\newtcolorbox{summarybox}[1]{ 
  colback=white,
  colframe=black,
  coltitle=black,
  colbacktitle=lightgray,
  width=0.99\columnwidth,
  boxrule=0.5pt,
  arc=2mm,
  title=\textbf{#1},
  center
}

\AtBeginDocument{%
  }

\setcopyright{acmlicensed}
\copyrightyear{2025}
\acmYear{2025}
\acmDOI{XXXXXXX.XXXXXXX}
\acmConference[Conference acronym 'XX]{Make sure to enter the correct
  conference title from your rights confirmation email}{June 03--05,
  2018}{Woodstock, NY}
\acmISBN{978-1-4503-XXXX-X/2018/06}

\RequirePackage[
  datamodel=acmdatamodel,
  style=acmnumeric,
  ]{biblatex}

\usepackage{longtable}
\begin{document}

\title{The Repeat Offenders: Characterizing and Predicting Extremely Bug-Prone Source Methods}


\author{Ethan Friesen}
\affiliation{%
  \institution{SQM Research Lab, Computer Science,}
  \institution{University of Manitoba}
  \city{Winnipeg}
  \country{Canada}}
\email{fries432@myumanitoba.ca}

\author{Sasha Morton-Salmon}
\affiliation{%
  \institution{SQM Research Lab, Computer Science,}
  \institution{University of Manitoba}
  \city{Winnipeg}
  \country{Canada}}
\email{mortonss@myumanitoba.ca}

\author{Md Nahidul Islam Opu}
\affiliation{%
  \institution{SQM Research Lab, Computer Science,}
  \institution{University of Manitoba}
  \city{Winnipeg}
  \country{Canada}}
\email{opumni@myumanitoba.ca}

\author{Shahidul Islam}
\affiliation{%
  \institution{SQM Research Lab, Computer Science,}
  \institution{University of Manitoba}
  \city{Winnipeg}
  \country{Canada}}
\email{islams32@myumanitoba.ca}

\author{Shaiful Chowdhury}
\affiliation{%
  \institution{SQM Research Lab, Computer Science,}
  \institution{University of Manitoba}
  \city{Winnipeg}
  \country{Canada}}
\email{shaiful.chowdhury@umanitoba.ca}
           

\begin{abstract}
Bug prediction has long been considered the ``prince'' of empirical software engineering research, and accordingly, a substantial body of work has focused on predicting bugs to enable early preventive actions. However, most existing studies operate at the class or file level, which practitioners have found to be of limited practical value. As a result, method-level bug prediction has gained increasing attention in recent years. Despite this shift, current method-level prediction models typically treat all buggy methods as equally fault-prone, regardless of whether a method has been associated with a bug once or repeatedly. We argue that methods involved in bugs multiple times—hereafter referred to as \emph{ExtremelyBuggy} methods—are more harmful than methods that are buggy only once. In this study, we investigate the prevalence of \emph{ExtremelyBuggy} methods, analyze their code quality metrics, and assess whether they can be predicted at the time of their introduction. In addition, we conduct a thematic analysis of 287 \emph{ExtremelyBuggy} methods to gain deeper insights into their characteristics.

Using a dataset of over 1.25 million methods extracted from 98 open-source Java projects, we find that only a small proportion of methods can be classified as \emph{ExtremelyBuggy}, yet these methods account for a disproportionately large share of bugs within a project. Although we observe statistically significant differences between \emph{ExtremelyBuggy} and other methods, \emph{ExtremelyBuggy} methods remain difficult to predict at their inception. Nevertheless, our manual analysis reveals recurring characteristics among these methods. These findings can help practitioners avoid harmful patterns in practice and guide future research toward developing features and models that better capture the unique properties of such methods.

\end{abstract}





\maketitle

\section{Introduction}

Software maintenance is one of the most expensive phases of the development lifecycle~\cite{borstler_role_2016}. A major contributor to this cost is the effort required to identify and correct software bugs, which alone can account for 50-70\% of the total development costs~\cite{wang2023bugpre, chowdhury_method-level_2024}. To no surprise, building models for accurate bug prediction is referred to as the \emph{prince of empirical software engineering research}~\cite{lanza2016tragedy} with researchers developing bug prediction models in an attempt to take early actions~\cite{chowdhury_method-level_2024, giger_method-level_2012, pascarella_performance_2020, rahman2017relationships, shivaji2012reducing, wang2023bugpre, wattanakriengkrai2020predicting}. The assumption is that if bug-prone code components can be identified early in the development lifecycle or prevented altogether, future maintenance costs will be significantly reduced~\cite{ibm}. 

Prior bug prediction models predominantly focused on the class or file levels~\cite{basili_validation_1996, gil_correlation_2017, zhou_ability_2010}. Unfortunately, these models remain underutilized in practice~\cite{tantithamthavorn2018experience}, at least partly because practitioners find it challenging to locate bugs at such coarse levels of granularity~\cite{chowdhury_method-level_2024, grund_codeshovel_2021, pascarella_performance_2020, lanza2016tragedy, shihab_industrial_2012,mondal_investigating_2019}. Practical use would require developers to inspect the entire flagged class or file to identify the specific location of the bug---a laborious and time-consuming task. Consequently, a significant amount of recent research has focused on bug prediction at the method-level granularity~\cite{mo2022exploratory, giger_method-level_2012,chowdhury_method-level_2024, Shippey, ferenc_automatically_2020, pascarella_performance_2020}. However, these models treat all buggy methods equally, without distinguishing between methods that were fixed once and those that required multiple bug fixes. 

In contrast to previous research, we focus on \emph{ExtremelyBuggy} methods---methods that have required bug fixes more than once---because identifying them early could prevent a disproportionate number of future failures and reduce the need for repeated maintenance on the same code fragments. Such methods may indicate architectural weak points or persistently complex areas that tend to accumulate defects over time. By identifying and targeting these methods, we aim to help practitioners prioritize their efforts where it matters most, enabling them to address a large number of bugs by focusing on a relatively small subset of problematic methods.

To support this goal, we analyze the characteristics of \emph{ExtremelyBuggy} methods in an effort to predict them at their inception---that is, as soon as they are pushed to a software project. To do so, we collected the change history of 1.25 million methods from 98 widely used open-source Java projects. To the best of our knowledge, this represents the largest and most comprehensive dataset to date focused on method-level bug prediction. Using this dataset, we address the following four research questions:

\textbf{RQ1:} What proportion of methods are \emph{ExtremelyBuggy}?

\textbf{Contribution 1:} We found that 0.04-6.63\% of methods are \emph{ExtremelyBuggy}. However, this small proportion of methods can often account for a large number of bugs in a project (even more than 90\% of the bugs in some projects). This implies that identifying this small fraction of methods early will significantly reduce future maintenance burden. 

\textbf{RQ2}: Do \emph{ExtremelyBuggy} methods exhibit signifantly different code quality than others?

\textbf{Contribution 2:} We found that, at their inception, \emph{ExtremelyBuggy} methods are significantly larger, less readable, and have lower maintainability scores compared to both \emph{Buggy} and \emph{NotBuggy} methods. This is encouraging, as it suggests that these methods exhibit distinguishable code quality that could be leveraged by machine learning models for early prediction. This observation led us to our next research question. 

\textbf{RQ3}: Can code quality-based machine learning models predict these \emph{ExtremelyBuggy} methods at their inception?

\textbf{Contribution 3:}  Unfortunately, our results indicate that machine learning algorithms perform poorly in distinguishing \emph{ExtremelyBuggy} methods from other methods. This is partly due to a significant class imbalance, as \emph{ExtremelyBuggy} methods represent only a small fraction of the overall dataset. Common techniques such as oversampling and undersampling did not lead to meaningful improvements in prediction performance. 

\textbf{RQ4}: What are the observable characteristics of \emph{ExtremelyBuggy} methods?

\textbf{Contribution 4:} The futility in predicting with machine learning models led us to manually investigate the characteristics of the \emph{ExtremelyBuggy} methods so that we can provide actionable guidance for the practitioners. As such, we conducted a thematic analysis~\cite{fereday2006demonstrating} on a curated dataset of 287 \emph{ExtremelyBuggy} methods. In general, we focused on three different dimensions while applying thematic analysis: visual issues, context of a method, and bug-fix type. Our analysis reveals some common themes that exist within these 287 methods. For example, we found that methods exhibiting self-admitted technical debt, implementing core logic or algorithms, or containing excessive exception handling or logging tend to become \emph{ExtremelyBuggy}. 


To help replication and extension in future method-level bug prediction research, we share our dataset publicly.\footnote{https://github.com/SQMLab/ExtremelyBuggyPublicData/tree/main/Dataset} 


\section{Related Work}

In this section, we discuss previous studies involving bug prediction and code metrics to show how they have motivated this paper.

\subsection{Software Maintenance and Bug Prediction}

Software maintenance is expensive~\cite{borstler_role_2016} and the bug-proneness of code components has been identified as one of the largest indicators of future maintenance burdens~\cite{chowdhury_good_2025}. In fact, reliably predicting software bugs is considered one of the holy grails~\cite{dambros_extensive_2010} and the \emph{prince} of empirical software engineering research~\cite{lanza2016tragedy}. As such, there has been an enormous effort ~\cite{bangash2020time,tantithamthavorn2018experience,d2010extensive, chowdhury_good_2025,rahman2017relationships,shivaji2012reducing,wang2023bugpre,wattanakriengkrai2020predicting,he2015empirical} by the research community over the last forty years~\cite{lenarduzzi2017analyzing} to predict bug-prone code components so that early actions can be taken.
Despite the great amount of research dedicated to predicting bugs, bug prediction models have yet to see widespread industry use~\cite{lanza2016tragedy, tantithamthavorn2018experience}. This is due, at least in part, to the fact that prior studies have predominantly focused on bug prediction at the file or class level~\cite{basili_validation_1996, gil_correlation_2017, zhou_ability_2010}, a granularity that practitioners often consider too coarse for practical use~\cite{mondal_investigating_2019, pascarella_performance_2020,grund_codeshovel_2021,chowdhury_method-level_2024}.

Although the most helpful models for practitioners would be to predict bugs at the line level, such models, however, do not offer high precision~\cite{wattanakriengkrai2020predicting}, as many lines can be similar just by chance~\cite{grund_codeshovel_2021}. Considering these factors, method-level bug prediction gained traction in recent years~\cite{ferenc_automatically_2020, giger_method-level_2012, hata2012bug, chowdhury_method-level_2024, pascarella_performance_2020, Shippey}. For example, Ferenc et al.~\cite{ferenc_automatically_2020} showed that accurate prediction models can be built by using different product and process metrics. Similar results were reported by Giger et al.~\cite{ giger_method-level_2012} and Hata et al.~\cite{hata2012bug}. Unfortunately, multiple recent studies~\cite{chowdhury_method-level_2024, pascarella_performance_2020} have shown that the high prediction accuracy reported by these studies is often inflated due to unrealistic evaluation scenarios. These studies have used k-fold cross-validation for model evaluation, which is prone to leakage between train and test data due to the time-sensitive nature of bug prediction data. And when tested on realistic scenarios, the accuracy of these models drops significantly~\cite{chowdhury_method-level_2024, pascarella_performance_2020}. 

Additionally, prior studies have not distinguished between methods involved in a single bug fix and those repeatedly involved in bug fixing. This distinction is critical because methods that frequently require maintenance—often referred to as maintenance hotspots—should be prioritized and allocated additional resources. Moreover, such methods may be more distinctive than ordinary bug-prone methods, a property that could be leveraged by prediction models to improve their accuracy. \emph{Motivated by this important gap in prior research, this paper focuses exclusively on source code methods that have been associated with bugs more than once. We investigate whether these methods exhibit distinct characteristics and whether they can be identified early in their lifecycle.}

\subsection{Code Metrics}
For bug prediction models to work, they need some quantifiable metrics that characterize the source code to use as inputs to the models. These metrics often come in the form of product metrics~\cite{ferenc_automatically_2020, mo2022exploratory, pascarella_performance_2020, chowdhury_revisiting_2022} and process metrics~\cite{graves2002predicting, mo2022exploratory, rahman2013and}. Process metrics are typically derived from the change history of a code component and have been shown to be effective predictors of bug-prone components~\cite{giger_method-level_2012, rahman2013and}. However, such metrics are not available immediately after a piece of code is written and therefore cannot support the early identification of maintenance-prone components. Waiting for sufficient change history to accumulate may allow these methods to cause significant issues, leading to high maintenance costs and potentially eroding stakeholder confidence.

Product metrics, on the other hand, are readily available and easy to compute at the inception of a code component. As a result, building maintenance prediction models using only product metrics has been a long-standing goal in software engineering research~\cite{romano_using_2011, pascarella_performance_2020, chowdhury_good_2025, nunez2017source}. Despite their extensive usage in software engineering research, some studies have found that product metrics, apart from \emph{size}, are not helpful for building models that predict future maintenance burdens~\cite{gil_correlation_2017, el2001confounding, sjoberg2012quantifying}. However, these studies with negative results were conducted at the class/file level, whereas studies that focused on method-level granularity have found product metrics as useful predictors of maintenance~\cite{landman2014empirical,chowdhury_method-level_2024, chowdhury_evidence_2025,chowdhury_revisiting_2022}. 

\emph{Motivated by the need for early prediction and the practicality of method-level source code metrics, we focus exclusively on these metrics to characterize and predict ExtremelyBuggy methods. Our goal is to predict these harmful methods as soon as they are pushed to a system.}

\section{Methodology}

\begin{figure}[h]
     \centering
     \begin{subfigure}[b]{\textwidth}
         \centering
         \includegraphics[width=1\textwidth, keepaspectratio]{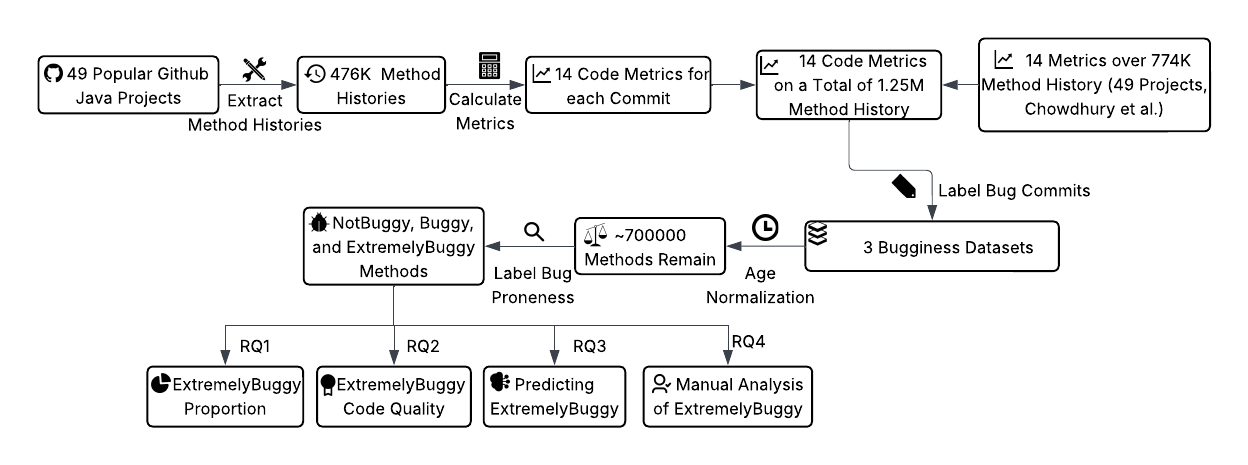}
     \end{subfigure}
        \caption{Methodology for collecting, preprocessing, labeling and analyzing our dataset.}
        \label{fig:MethodologyFlow}
\end{figure}

Similar to previous method-level bug prediction studies~\cite{pascarella_performance_2020, chowdhury_method-level_2024, mo2022exploratory, Shippey}, we focused on Java projects only. This also mitigates threats associated with code metric–based studies that rely on projects written in multiple programming languages~\cite{Zhang:2013}. In this section, we outline the overall methodology, as presented in Figure~\ref{fig:MethodologyFlow}, including project selection, change history tracking, and the identification of bug-proneness indicators. We also detail the data preprocessing steps and the selection of code metrics. To improve readability, methodologies specific to individual research questions are presented alongside each corresponding research question. 

\subsection{Dataset}
We need to build a dataset of methods along with their change histories, so that we can label each method’s bug-proneness by identifying bug- and fix-related keywords in all the commit messages associated with that method. A substantial portion of our dataset originates from the work of Chowdhury et al.~\cite{chowdhury_method-level_2024}, which comprises 774{,}051 Java methods drawn from 49 open-source projects. Each method in this dataset is accompanied by its complete change history and is enriched with metadata such as commit messages, timestamps, authorship information, and change types. The dataset additionally includes code diffs for all historical modifications, binary labels indicating whether each change corresponds to a bug fix, and the number of methods modified within the associated commit. These 49 projects have been widely utilized, either fully or partially, in prior method-level software engineering studies~\cite{ray_naturalness_2016,gil_correlation_2017,Palomba:2017,spadini_relation_2018,grund_codeshovel_2021,chowdhury_evidence_2025,chowdhury_method-level_2024,chowdhury_good_2025}. Since \emph{ExtremelyBuggy} methods are even rarer than buggy methods (as shown later), we aimed to expand the dataset of Chowdhury et al. Therefore, we collected additional open-source projects and extracted their method-level change histories.

\subsubsection{Project Selection}
We employed the GitHub REST API\footnote{\url{https://docs.github.com/en/rest}}
 and the PyGitHub library\footnote{\url{https://github.com/PyGithub/PyGithub}}
 to collect open-source projects. To ensure the quality and reliability of our dataset, we followed recommended best practices for mining GitHub data~\cite{kalliamvakou2014promises}. We began by retrieving the 1000 most-starred GitHub repositories that satisfied several additional criteria. First, to reflect contemporary industry practices, we restricted our search to repositories created within the past 15 years\footnote{Created after January 1st, 2010} and excluded repositories that had been inactive for more than one year.\footnote{No commits since May 13, 2024} Second, to ensure representativeness of real-world Java projects, we excluded templates, archived repositories, and forks by specifying appropriate parameters in our GitHub API queries. Third, we required repositories to contain at least $2000$ commits to guarantee substantial change history. We further filtered out repositories whose codebase consisted of less than $95\%$ Java, thereby ensuring the availability of a sufficiently large number of Java methods for extraction.

Following the automated filtering process, we manually reviewed all remaining repositories to verify adherence to the above criteria and to confirm that their descriptions were written in English. Ensuring English descriptions facilitates the identification of bug-fixing commits using English-language keywords and supports subsequent manual qualitative analysis of commit messages. After this review process, we selected 49 additional projects from the available projects. Combined with the 49 projects from Chowdhury et al.~\cite{chowdhury_method-level_2024}, our final dataset comprises 98 projects in total. The complete list of filtered projects is included in our repository.\footnote{\url{https://github.com/SQMLab/ExtremelyBuggyPublicData/tree/main/AdditionalFigures/RepositoryList.md}} 

\subsubsection{Extracting Method-level History}
To quantify the bug-proneness of a method, it is necessary to determine how frequently that method has been involved in bug-fixing activities, which in turn requires capturing its complete change history. To obtain this historical information, similar to Chowdhury et al.~\cite{chowdhury_method-level_2024}, we employed the CodeShovel tool~\cite{grund_codeshovel_2021}. CodeShovel has demonstrated high accuracy in identifying method-level change histories and is robust to complex code transformations, including method relocations across files~\cite{grund_codeshovel_2021}. Its reliability has also been independently evaluated by industry practitioners, who reported similarly strong accuracy~\cite{grund_codeshovel_2021}. Although a more recent tool, CodeTracker~\cite{jodavi_accurate_2022}, has been shown to achieve higher accuracy than CodeShovel, its substantially slower performance made it challenging to use it for tracing the histories of the large number of methods we have in our dataset. Moreover, CodeTracker has not undergone independent validation by industry professionals. 

To support large-scale and reliable analysis, we gathered structured JSON outputs for approximately 1.25 million methods across 98 projects, capturing their complete method-level evolution across commits. These outputs provide a robust foundation from which we extract both code metrics and change metrics for every method revision included in our dataset.
To the best of our knowledge, no prior bug prediction study has included such a large number of projects and methods. We believe this dataset can serve as a valuable resource for future research on method-level bug prediction.

\subsection{Code Metrics}
Code metrics have traditionally served as indicators of quality and maintainability of code~\cite{gil_correlation_2017, pascarella_performance_2020, chowdhury_method-level_2024}. Encouraged by previous research in method-level code metrics~\cite{mo2022exploratory,pascarella_performance_2020, chowdhury_evidence_2025}, we used 14 code metrics to determine if the code quality indicators of the \emph{ExtremelyBuggy} methods are different than others. These metrics were calculated using the same tool used in ~\cite{chowdhury_evidence_2025,chowdhury_method-level_2024, chowdhury_revisiting_2022}. The selected metrics are described as follows. 

\textbf{\emph{Size.}} Size is one of the simplest metrics, yet it is considered one of the most important with regard to software quality and maintenance~\cite{gil_correlation_2017, el2001confounding, chowdhury2022empirical}. We used the number of source lines of code without comments and blank lines as our size measurement (referred to as SLOCStandard)~\cite{chowdhury_method-level_2024, landman2014empirical, ralph2018construct, chowdhury_evidence_2025}.

\textbf{\emph{Readability.}} The ability to read and understand existing code is an important part of software maintenance~\cite{scalabrino2016improving}. Code readability can be broadly defined by how easy it is for a developer to understand the structure and flow of source code~\cite{buse2009learning, posnett2011simpler, borstler_role_2016, johnson2019empirical}. Developers can more easily understand and therefore modify readable code with reduced risk of issues. We used two distinct readability metrics in this paper: \textit{ Buse et al.}~\cite{buse2009learning} (\emph{Readability}) and \textit{Posnett et al.}~\cite{posnett2011simpler} (\emph{SimpleReadability}). Both provide a value for each method in the range $[0,1]$.

\textbf{\emph{Complexity and Testability.}} We used the McCabe metric~\cite{mccabe1976complexity} to analyze the complexity of methods. This metric represents the number of independent paths in a method; the greater the number of paths, the more challenging it becomes to test the method thoroughly. In addition, we used the number of control variables (\emph{NVAR}), the number of comparisons (\emph{NCOMP}), and the McClure metric~\cite{mcclure1978model}, which is calculated as the sum of \emph{NVAR} and \emph{NCOMP}. These metrics offer additional insights into method complexity that are not captured by the McCabe metric. Hindle \textit{et al.}~\cite{hindle2008reading} proposed the proxy indentation metric (\emph{IndentSTD}~\cite{chowdhury_revisiting_2022}), which performs similarly to the McCabe metric in capturing code complexity but, unlike McCabe, does not require a language-specific parser. We included \emph{IndentSTD} in our metric set as well. Finally, we used the \emph{MaximumBlockDepth} metric, which measures the maximum nested block depth, as deep nesting within a method can further complicate testing.

\textbf{\emph{Dependency.}} When methods are highly coupled, a bug within one method is more likely to affect another. To measure this dependency between methods, we used \emph{totalFanOut}, which is the total number of methods called by a given method.

\textbf{\emph{Maintainability.}} \emph{Maintainability Index} (MI)~\cite{oman1992metrics} is a composite software quality metric that integrates \emph{HalsteadVolume}, which quantifies the amount of information a reader must process to comprehend the code’s functionality, along with \emph{McCabe} complexity and \emph{Size}. The MI is computed using the following formula:

\begin{equation}
MI = 171 - 5.2 \times \ln(HalsteadVolume) - 0.23 \times (McCabe) - 16.2 \times \ln(Size)
\end{equation}

It produces a single score $(-\infty,171]$, designed to reflect the maintainability of a given method, where a higher score reflects better maintainability. Industry standard tools such as Visual Studio\footnote{\url{https://learn.microsoft.com/en-us/visualstudio/code-quality/code-metrics-maintainability-index-range-and-meaning?view=vs-2022}} use this metric, making it a worthy inclusion for our purposes.

\textbf{\emph{Other.}} The number of parameters (\emph{Parameters}), the number of local variables (\emph{LocalVariables}), and the Halstead Length (\emph{Length}), which measures the total number of operators and operands, are also included alongside the other metrics. We also intended to include the \emph{FanIn} metric in our analysis, but were unable to do so due to the complexity involved. For example, consider the Hadoop project, which contains approximately 70,000 methods. Since our goal is to make predictions at method inception, we need to capture code metrics at the time each method was first introduced. Measuring the \emph{FanIn} metric would require constructing a separate call graph for the whole codebase at each method's introduction commit. Unfortunately, building such call graphs is extremely time-consuming, as it requires preprocessing the entire codebase. Given that we work with 98 projects—each potentially requiring thousands of unique call graphs corresponding to thousands of distinct commits—this approach was computationally infeasible within the scope of our study. This limitation was mentioned in other similar studies as well~\cite{chowdhury_method-level_2024, chowdhury_evidence_2025}.


\subsection{Labeling Bug-fix Commits}
The bug-proneness of a method is historically estimated by extracting bug-related keywords from commit messages. A commonly used set of such keywords (and their variations) includes: \textit{error, bug, fix, issue, mistake, incorrect, fault, defect, and flaw}~\cite{ray_naturalness_2016, mockus_identifying_2000}. However, this keyword set often produces false positives, primarily due to the inclusion of the keyword \textit{issue}~\cite{chowdhury_method-level_2024}. To reduce these false positives, Chowdhury \textit{et al.}~\cite{chowdhury_method-level_2024} proposed removing the keyword \textit{issue} and focusing on commit messages that contain both bug-related and fix-related keywords. Although this method significantly improves the accuracy of identifying bug-fix commits, it does not eliminate noise caused by tangled commits—commits that include changes to both bug-related and unrelated methods—leading to incorrectly labeled methods.

While handling tangled code changes remains an active research area~\cite{herbold2022fine, opu2025llm, chowdhury_method-level_2024}, we conduct our analysis using three distinct datasets to mitigate the impact of tangled commits. The underlying hypothesis is that if similar results are observed across all three datasets, the findings are more robust and less likely to be impacted by tangled changes~\cite{chowdhury_evidence_2025}.

\begin{enumerate}
\item \textbf{HighRecall Dataset.} Here, we used all the above-mentioned keywords and their variations (include \textit{issue}) and did not consider the problem of tangled changes. While this approach obtains high recall since any change remotely related to a bug-fix is labeled as a bug-fix, it results in low precision as many changes not related to a bug-fix are labeled as such. Since this dataset is unlikely to miss any bugs, it is well-suited for estimating the proportion of bugs that can be identified by considering only the \emph{ExtremelyBuggy} methods (RQ1). 
\item \textbf{HighPrecision Dataset.} In this dataset, we used the set of keywords introduced by Chowdhury \textit{et al.}~\cite{chowdhury_method-level_2024}, which aims at increasing the precision of bug-fix labeling. To remove the effects of tangled changes, we only labeled a change as a bug-fix if exactly one method was changed in that commit---if there is only one modified method in a bug-fix commit, that method was certainly a bug-prone method~\cite{chowdhury_method-level_2024}. While this approach achieves a very high level of precision, it ultimately reduces recall, as many bug fixes may be ignored. Because this dataset is small and precise, it is particularly well suited for the thematic analysis (RQ4) aimed at understanding the properties of \emph{ExtremelyBuggy} methods.
\item \textbf{Balanced Dataset.} We used the same set of keywords as the HighPrecision Dataset; however, we allowed up to 5 methods changed in a commit for it to be labeled as a bug fix. This approach aims at achieving a balance between precision and recall.
\end{enumerate}

\subsection{Age Normalization}
\begin{figure*}[h]
     \centering
     \begin{subfigure}[b]{0.4\textwidth}
         \centering
         \includegraphics[width=\textwidth]{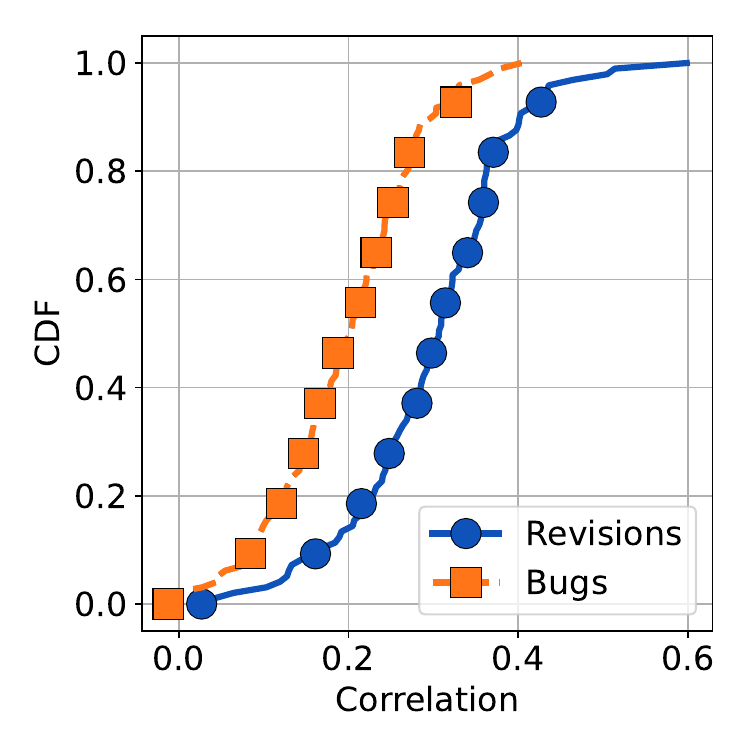}
         \caption{Cumulative Distribution Function (CDF) of correlation between age and both revisions and bug-proneness for all 98 projects}
         \label{fig:AgeCorrelation}
     \end{subfigure}
     \hspace{0.5cm}
     \begin{subfigure}[b]{0.42\textwidth}
         \centering
         \includegraphics[width=\textwidth]{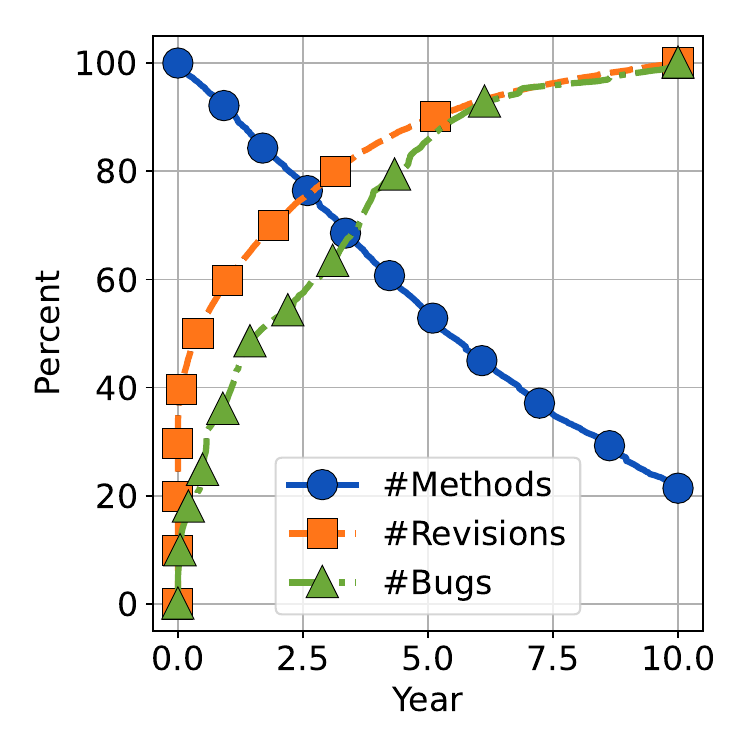}
         \caption{Percent of total methods, revisions, and bugs retained at different age-normalization thresholds.}
         \label{fig:AgeNormThresholds}
     \end{subfigure}
        \caption{Reason for selecting a 5-year threshold for age normalization. To enhance readability, we only use 11 markers when presenting each distribution; however, all data points are represented.}
        \Description{Both figures use the HighRecall Dataset to measure bugs}
        \label{fig:AgeNormalization}
\end{figure*}

The age of the methods in our dataset varies significantly. This means that comparing newer methods to older methods may introduce bias, as older methods have more time to undergo changes and hence, undergo bug-fix related changes. This assumption is supported by Figure~\ref{fig:AgeCorrelation}, which shows that for many projects there is a positive correlation—measured using Kendall’s $\tau$ correlation coefficient with $p \le 0.05$—between age and both revisions and bugs. Specifically, for approximately 60\% of projects, the correlation between a method's age and its number of bugs is $\ge 0.2$. While weak, such correlations should still be considered to minimize potential confounding in our analysis.

Due to this correlation, we removed all methods that are under 5 years old from our dataset. However, we still cannot directly compare methods that are 5 years old with those that are more than 5 years old for the aforementioned reasons. Thus, we also remove all changes made after the first 5 years from the remaining methods. This, unfortunately, resulted in us losing 593,409 methods and all methods from the following 6 projects from our dataset: \textit{'junit5', 'SmartTube', 'RxJava', 'Essentials', 'titan', 'xpipe'}. Since we are mainly interested in \emph{ExtremelyBuggy} methods, we care more about retaining change history---allowing methods to exhibit their true bug-proneness. Thus, the reasoning behind a 5-year threshold can be summarized in Figure~\ref{fig:AgeNormThresholds}. Here we see that 5 years allows us to retain \textasciitilde90\% of changes and \textasciitilde85\% of bugs while still saving \textasciitilde50\% of the methods in our dataset. If we decrease the threshold, we lose more bug-proneness information in our dataset, which subsequently reduces the number of \emph{ExtremelyBuggy} methods. 

The end result of this age-normalization is a dataset of 691,610 methods from 92 projects, where each method has exactly 5 years of change history. All the projects in our dataset, after age normalization, as well as the number of \emph{ExtremelyBuggy} methods in each project for the Balanced dataset can be viewed in our public repository.\footnote{\url{https://github.com/SQMLab/ExtremelyBuggyPublicData/blob/main/AdditionalFigures/AgeNormalizedRepositoryList.md}}

\subsection{Labeling Bug Proneness}
For all three datasets, we assigned each method to one of three categories based on the number of bug-fix commits in which it was involved:
\begin{itemize}
\item \emph{NotBuggy}: Methods that were never involved in a bug-fix commit.
\item \emph{Buggy}: Methods that were involved in exactly one bug-fix commit.
\item \emph{ExtremelyBuggy}: Methods that were involved in more than one bug-fix commit.
\end{itemize}


\subsection{Statistical Tests}
After evaluating our data with the Anderson-Darling normality test~\cite{razali2011power}, we found that most of the distributions in our data (e.g., the code metrics) do not follow a normal distribution. Therefore, we adopted the non-parametric Wilcoxon rank sum test to assess if there is a statistical difference between two given distributions. To understand the size of those difference, we employed the non-parametric Cliff's \textit{d} and categorized the effect sizes following Hess et al.~\cite{hess2004robust}: \textit{Negligible (N)} ($\delta < 0.147$), \textit{Small (S)} ($0.147 \leq \delta < 0.33$), \textit{Medium (M)} ($0.33 \leq \delta < 0.474$), and \textit{Large (L)} ($\delta \geq 0.474$).  Both of these statistical tests do not assume normally distributed distributions and have been widely used in software engineering research~\cite{he2015empirical, chen2020savior, pecorelli2020testing, chowdhury2019greenscaler, bangash2020time}. In addition, we used the non-parametric Kendall's $\tau$ correlation coefficient instead of Pearson's \textit{r}~\cite{inozemtseva2014coverage, gil_correlation_2017, chowdhury2019greenscaler}. Unless otherwise stated, we use a significance threshold of $p \le 0.05$ when evaluating the statistical significance of correlation coefficients.

\section{Approach, Analysis and Results}
In this section, we discuss the approach of each RQ along with the findings. 

\subsection{RQ1: Proportion of \emph{ExtremelyBuggy} Methods and Their Impact}

Past research has shown that a high proportion of changes comes from a small segment of code~\cite{grosfeld2007pareto, chowdhury_good_2025}. Motivated by this observation, we investigate what proportion of methods are \emph{ExtremelyBuggy} and how many bugs the \emph{ExtremelyBuggy} methods account for in a project.

\begin{table}[h]
    \centering
    \caption{Number of \emph{Buggy} and \emph{ExtremelyBuggy} methods in all 3 datasets.}
    \begin{tabular}{lrr}
    \toprule
         \textbf{Dataset} & \textbf{\#\emph{Buggy} (\%)} & \textbf{\#\emph{ExtremelyBuggy} (\%)}  \\
    \midrule
         HighRecall & 79715 (11.53) & 45860 (6.63) \\
         HighPrecision & 2704 (0.39) & 287 (0.04) \\
         Balanced & 8024 (1.16) & 1195 (0.17) \\
    \bottomrule
    \end{tabular}
    \label{tab:EBMethodCount}
\end{table}

Table~\ref{tab:EBMethodCount} shows the number of \emph{Buggy} and \emph{ExtremelyBuggy} methods for all three different datasets when all 92 projects are aggregated together. Although the high recall dataset contains a large number of \emph{ExtremelyBuggy} methods (6.63\%), these numbers decline significantly when we observe the HighPrecision and Balanced datasets. Specifically, only 0.04\% (or 287 of the \textasciitilde700000) methods are \emph{ExtremelyBuggy} in the HighPrecision dataset. Another observation is that, compared to the HighRecall dataset, the ratio of \emph{Buggy} to \emph{ExtremelyBuggy} methods is smaller in the HighPrecision and balanced datasets. In the HighRecall dataset, approximately 37\% of all methods with at least one bug-fix (\emph{Buggy} and \emph{ExtremelyBuggy}) are \emph{ExtremelyBuggy} whereas in the HighPrecision and Balanced datasets, this percentage drops to approximately 10\% and 13\%, respectively. 

\begin{figure}[h]
     \centering
     \begin{subfigure}[b]{0.32\textwidth}
         \centering
         \includegraphics[width=\textwidth]{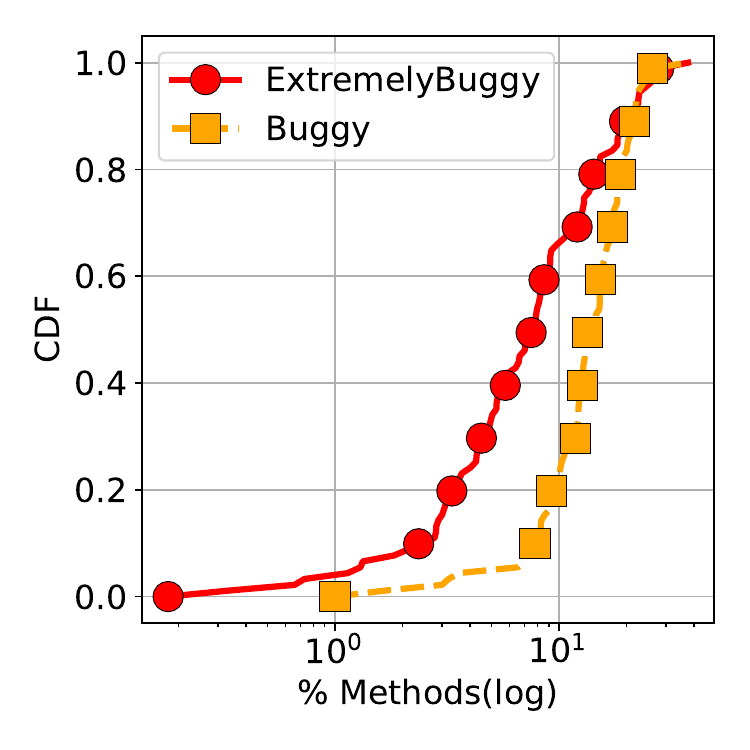}
         \caption{HighRecall Dataset}
         \label{fig:RecallCounts}
     \end{subfigure}
     \hfill
     \begin{subfigure}[b]{0.32\textwidth}
         \centering
         \includegraphics[width=\textwidth]{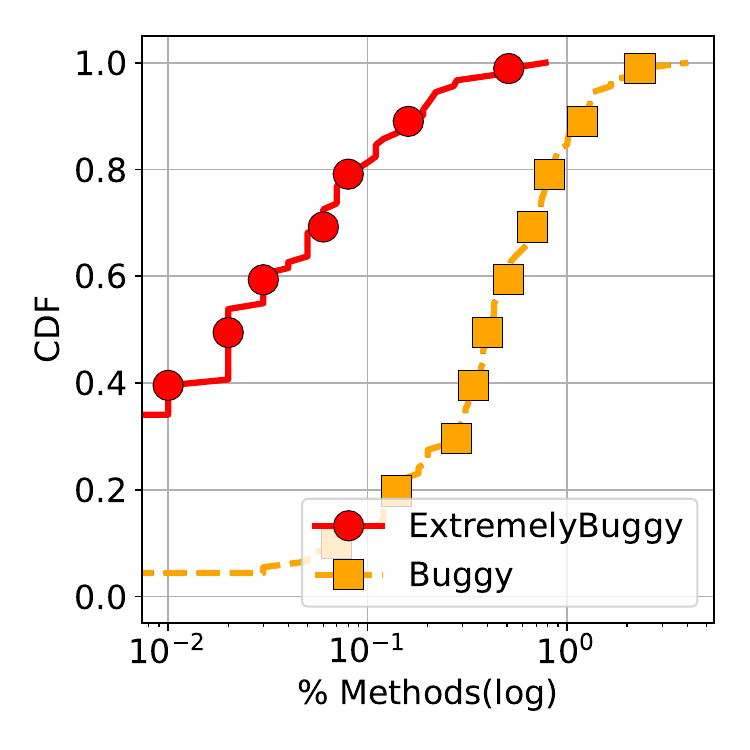}
         \caption{HighPrecision Dataset}
         \label{fig:PrecisionCounts}
     \end{subfigure}
     \hfill
     \begin{subfigure}[b]{0.32\textwidth}
         \centering
         \includegraphics[width=\textwidth]{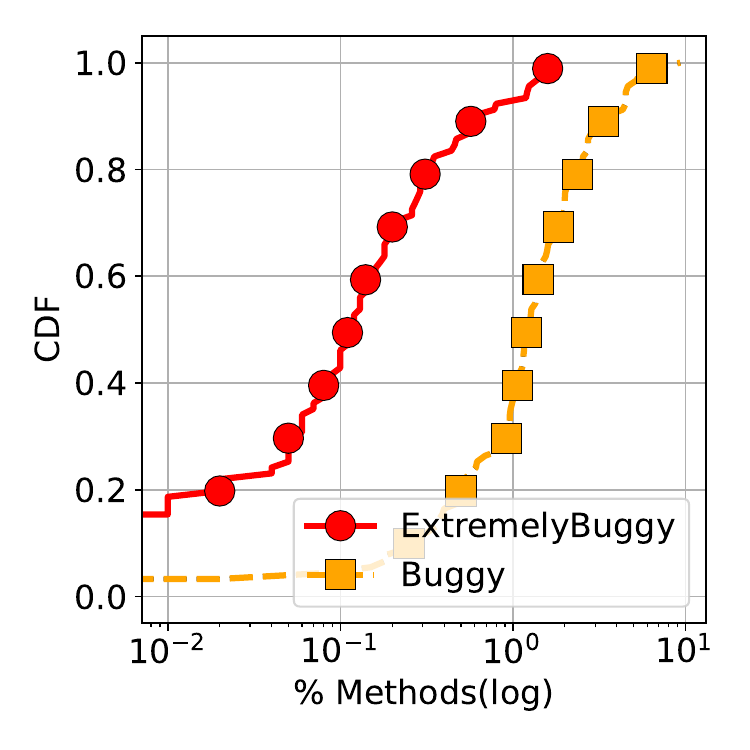}
         \caption{Balanced Dataset}
         \label{fig:BalancedCounts}
     \end{subfigure}
        \caption{CDF to show the percentages of methods for each project that are \emph{Buggy} and \emph{ExtremelyBuggy}. To enhance readability, we only use 11 markers when presenting each distribution; however, all 92 projects are represented.}
        \Description{DFs of the percent of methods in that are \emph{Buggy} and \emph{ExtremelyBuggy}}
        \label{fig:EBMethodCounts}
\end{figure}

Figure~\ref{fig:EBMethodCounts} presents the results at the project level. For example, in the HighRecall dataset, more than 10\% of methods are buggy in 80\% of the projects, and more than 10\% of methods are \emph{ExtremelyBuggy} in 40\% of the projects. As expected, these percentages decrease substantially in the Balanced dataset and even more markedly in the HighPrecision dataset. In both datasets, less than 1\% of the methods are buggy in most projects, which is even lower for the \emph{ExtremelyBuggy} methods. These findings complement our earlier aggregated-level findings presented in Table~\ref{tab:EBMethodCount}. We also found that, for the HighPrecision dataset, \textasciitilde35\% of projects have no \emph{ExtremelyBuggy} methods. 

\begin{figure}[h]
     \centering
     \begin{subfigure}[b]{0.32\textwidth}
         \centering
         \includegraphics[width=\textwidth]{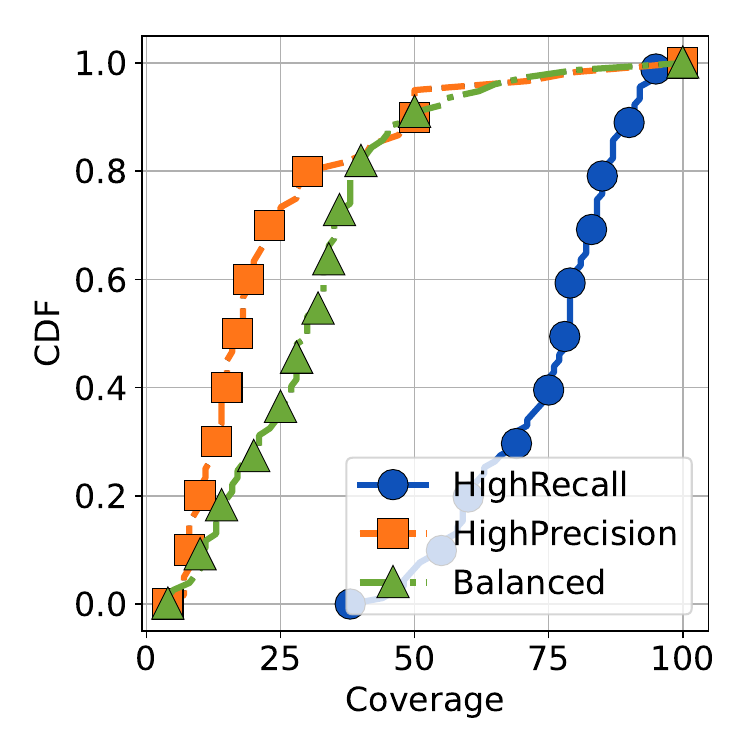}
     \end{subfigure}
        \caption{CDFs of the percent of bugs covered by the \emph{ExtremelyBuggy} methods for each of the three datasets for all 92 projects, excluding projects with no \emph{ExtremelyBuggy} methods. To enhance readability, we only use 11 markers when presenting each distribution.}
        \Description{CDFs of the percent of bugs covered by the \emph{ExtremelyBuggy} methods }
        \label{fig:ExtremelyBuggyCoverage}
\end{figure}

We now investigate what percent of bugs are captured by the \emph{ExtremelyBuggy} methods alone. We counted the number of bugs in the \emph{ExtremelyBuggy} methods in each project and divided it by the total number of bugs in the project for each of the 92 projects. To avoid overcounting the number of bugs in each project, we only count a bug once for each unique commit. The results for each dataset are presented in figure~\ref{fig:ExtremelyBuggyCoverage}. For the HighRecall dataset—the most suitable dataset for this analysis, as it is likely to have fewer false negatives than the other two datasets—we observe that \emph{ExtremelyBuggy} methods account for a large proportion of the bugs. For \textasciitilde80\% of projects, the \emph{ExtremelyBuggy} methods accounted for $\ge60\%$ of bugs, and for \textasciitilde50\% of projects, the \emph{ExtremelyBuggy} methods covered $\ge75\%$ of bugs. If we consider the percentage of \emph{ExtremelyBuggy} methods (Table~\ref{tab:EBMethodCount}), this implies that, on average, only 6–7\% of methods in many projects account for more than 75\% of all bugs in those projects. 

Not surprisingly, the results are much different for the HighPrecision and Balanced datasets. For these two datasets, we see much less bug coverage of the top percent of methods with bugs. However, this is expected as in the HighPrecision and Balanced datasets, many of the \emph{ExtremelyBuggy} methods and the number of bugs they covered were excluded for maintaining higher precision.


\begin{summarybox}{RQ1 Summary}
\emph{ExtremelyBuggy} methods are few in number yet account for a large share of total bugs. This is encouraging because by focusing on a small fraction of the codebase, practitioners can substantially reduce future maintenance effort. This also underscores the importance of early prediction and identification of these methods.
\end{summarybox}
\subsection{RQ2: Code Quality of \emph{ExtremelyBuggy} Methods}
To answer this research question, we analyze how the code quality indicators of \emph{ExtremelyBuggy} methods differ from those of both \emph{Buggy} and \emph{NotBuggy} methods. Since our goal is to identify \emph{ExtremelyBuggy} methods at the time of their introduction, we compare the code metrics of all three categories—\emph{NotBuggy}, \emph{Buggy}, and \emph{ExtremelyBuggy}—at the point when the methods are first added to the codebase.

\begin{figure}[h]
     \centering
     \begin{subfigure}[b]{0.33\textwidth}
         \centering
         \includegraphics[width=\textwidth]{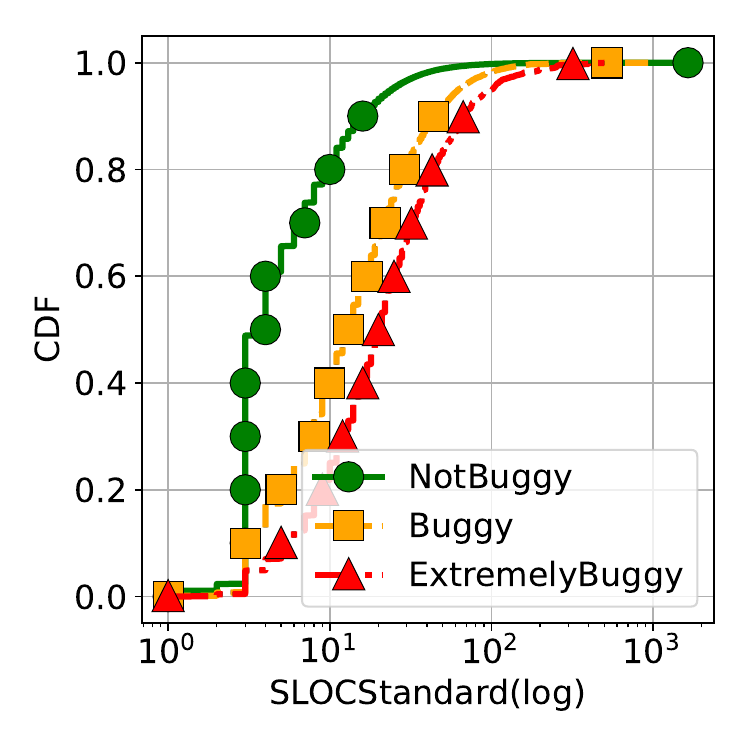}
         \label{fig:SLOC-differences}
     \end{subfigure}
     \hfill
     \begin{subfigure}[b]{0.33\textwidth}
         \centering
         \includegraphics[width=\textwidth]{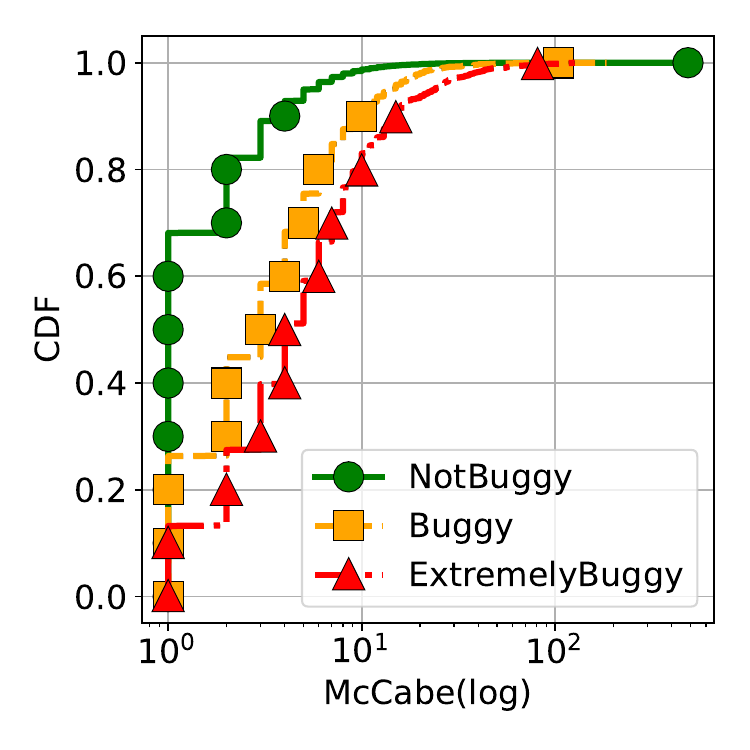}
         \label{fig:McCabe-differences}
     \end{subfigure}
     \hfill
     \begin{subfigure}[b]{0.33\textwidth}
         \centering
         \includegraphics[width=\textwidth]{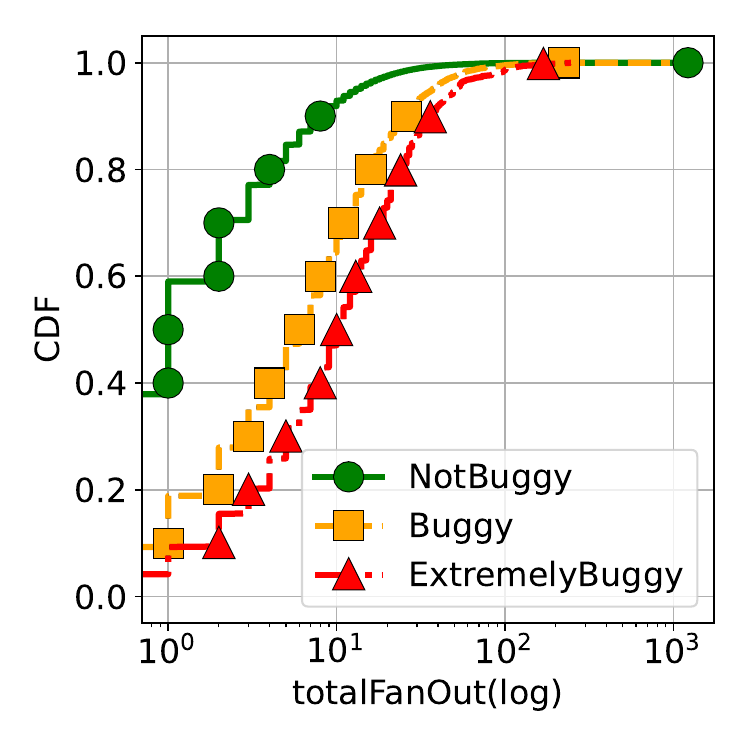}
         \label{fig:Fanout-differences}
     \end{subfigure}
     \hfill
     \begin{subfigure}[b]{0.33\textwidth}
         \centering
         \includegraphics[width=\textwidth]{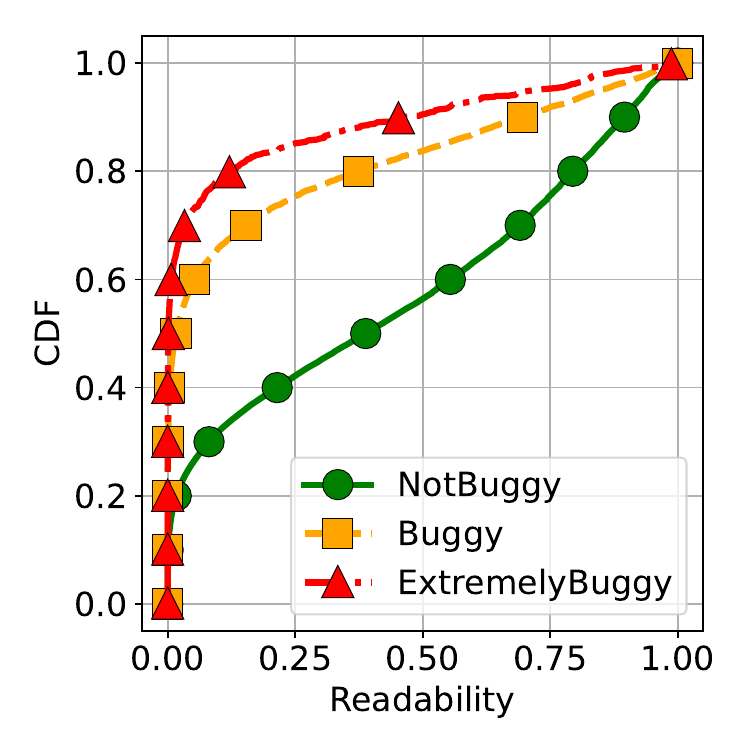}
         \label{fig:Readability-differences}
    \end{subfigure}
    \hfill
         \begin{subfigure}[b]{0.33\textwidth}
         \centering
         \includegraphics[width=\textwidth]{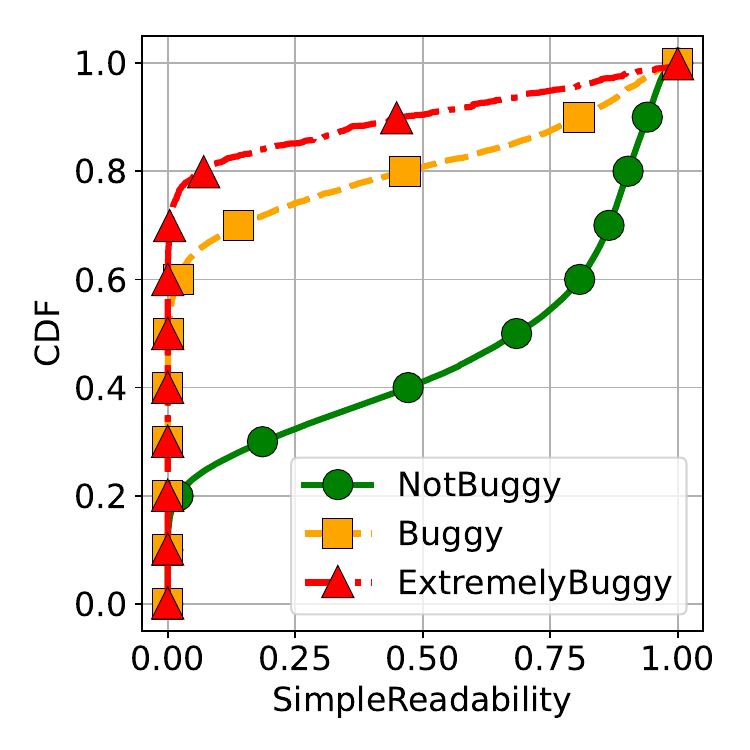}
         \label{fig:SimpReadability-differences}
     \end{subfigure}
     \hfill
     \begin{subfigure}[b]{0.33\textwidth}
         \centering
         \includegraphics[width=\textwidth]{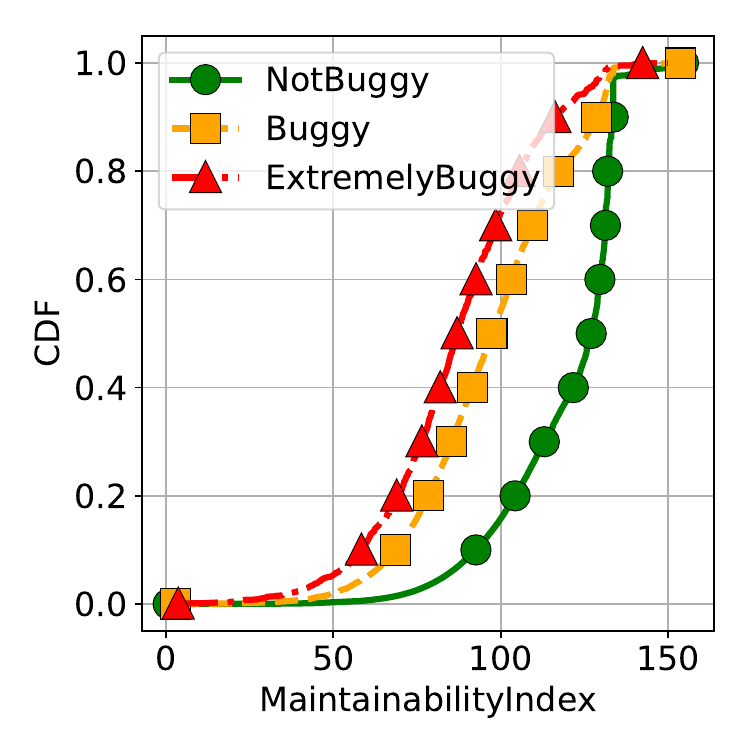}
         \label{fig:MaintIndex-differences}
     \end{subfigure}
        \caption{CDFs showing the distributions of different code metrics at the first commit between \emph{NotBuggy}, \emph{Buggy}, and \emph{ExtremelyBuggy} methods for the Balanced dataset. Note that 34 of the \textasciitilde700000 methods had a \emph{MaintainabilityIndex} below zero. To improve readability, we excluded these methods when constructing the \emph{MaintainabilityIndex} graph.}
        \Description{CDFs showing the distributions of different code metrics at the first commit}
        \label{fig:CodeMetricDifs}
\end{figure}

Figure~\ref{fig:CodeMetricDifs} compares the three different method classes for six of the selected code metrics using the Balanced dataset; the results are similar for all code metrics across all three datasets. Clearly, \emph{ExtremelyBuggy} methods are larger, less readable, more complex, and have higher dependencies than both the \emph{Buggy} and \emph{NotBuggy} methods. 


To confirm these visual findings, we also performed a statistical analysis to compare \emph{ExtremelyBuggy} methods with both \emph{Buggy} and \emph{NotBuggy} methods. In Table~\ref{tab:AggregatedAnalysisCodeMetrics}, we present the results when comparing the \emph{ExtremelyBuggy} methods to \emph{Buggy} and \emph{NotBuggy} methods grouped together for the Balanced dataset. The comparisons were done on all 92 projects' distributions aggregated together. We can see that all the P-values are significant ($P \le 0.05$) and that all the effect sizes are \textit{large}, except for \textit{Parameters}, which is \textit{medium}. This indicates that there is a substantial difference between \emph{ExtremelyBuggy} methods and all other methods at inception. Additionally, the sign of the effect size is negative ($-$) for \textit{Readability}, \textit{SimpleReadability}, and \textit{MaintainabilityIndex}, indicating that \emph{ExtremelyBuggy} methods are less readable and less maintainable. For other code metrics, the sign is positive ($+$), indicating that the \emph{ExtremelyBuggy} methods are larger and more complex. The results are very similar for the HighPrecision dataset. However, the effect sizes for the HighRecall dataset are mostly \textit{medium}. This is because the HighRecall dataset has a higher number of false positives when identifying the \emph{ExtremelyBuggy} methods, resulting in more noise and less distinction of the \emph{ExtremelyBuggy} methods.

\begin{table}[ht]
    \centering
    \caption{Results of the statistical significance test (Wilcoxon rank sum test) and effect size (Cliff's delta) comparing \emph{ExtremelyBuggy} methods with the combined group of \textit{Buggy} and \textit{NotBuggy} methods for the Balanced dataset. The method groups used in the comparison are formed by aggregating methods from all projects.}
    \begin{tabular}{lccc}
        \toprule
            \textbf{Metric} & \textbf{P-value} & \textbf{Sign(+/-)} & \textbf{Effect Size} \\
        \midrule
            SLOCStandard & 0.00 & + & large \\
            Readability & 0.00 & - & large \\
            SimpleReadability & 0.00 & - & large \\
            NVAR & 0.00 & + & large \\
            NCOMP & 0.00 & + & large \\
            Mcclure & 0.00 & + & large \\
            McCabe & 0.00 & + & large \\
            IndentSTD & 0.00 & + & large \\
            MaximumBlockDepth & 0.00 & + & large \\
            totalFanOut & 0.00 & + & large \\
            Length & 0.00 & + & large \\
            MaintainabilityIndex & 0.00 & - & large \\
            Parameters & 0.00 & + & medium \\
            LocalVariables & 0.00 & + & large \\
        \bottomrule
    \end{tabular}

    \label{tab:AggregatedAnalysisCodeMetrics}
\end{table}

In addition to the aggregated analysis, we also perform individual project analysis to obtain generalizable observations.  Table~\ref{tab:IndividualProjectAnalysisCodeMetrics} presents the results for the \emph{ExtremelyBuggy} compared to the \emph{Buggy} and \emph{NotBuggy} methods grouped together for the Balanced dataset. For example, for 78.21\% of projects, the difference in \textit{SLOCStandard} between \emph{ExtremelyBuggy} and other methods is statistically significant ($P \le 0.05$). For these projects, the difference was \textit{large} (L) for 98.36\% of the projects, and for the remaining 1.64\% it was \textit{medium} (M). For all code metrics, some projects (usually approximately 20-30\%) did not show statistical differences between \emph{ExtremelyBuggy} and other methods. This is because some projects in the Balanced dataset have a very low number of \emph{ExtremelyBuggy} methods, making it difficult to discern differences in distributions. However, for the projects that did have a statistical difference, that difference was almost always \textit{large}. Out of the 14 code metrics, only one, \textit{Parameter}, exhibited distinct characteristics. While more than 62\% of projects consistently demonstrated statistical significance for all other metrics, \textit{Parameter} differed for only 20\% of projects, indicating that parameter count is not as influential as other metrics for identifying \emph{ExtremelyBuggy} methods. 

For the HighPrecision dataset, approximately 40\% of projects exhibited statistical differences between \emph{ExtremelyBuggy} and other methods for all metrics, with the effect size remaining almost always large. For the HighRecall dataset, approximately 90-95\% of projects demonstrated statistical differences for all metrics; however, the effect size ranged from small (S) to large (L) (typically \textasciitilde30\% in each category). This is again due to the lower precision in the HighRecall dataset.

\begin{table}[ht]
    \centering
    \caption{The results of statistical significance test (Wilcoxon rank sum test) and effect size (Cliff's delta) comparing \emph{ExtremelyBuggy} methods with the combined group of \textit{Buggy} and \textit{NotBuggy} methods for the Balanced dataset, evaluated per project. Each value denotes the percentage of projects exhibiting the corresponding behavior. For instance, for the \textit{totalFanOut} metric, 80.77\% of projects show a statistically significant difference, and in 96.83\% of those cases, the effect size is classified as \textit{large} (L).}
    \begin{tabular}{lccccc}
        \toprule
            \textbf{Metric} & \textbf{$P \le 0.05$} & \textbf{N} & \textbf{S} & \textbf{M} & \textbf{L} \\
        \midrule
            SLOCStandard & 78.21 & 0.00 & 0.00 & 1.64 & 98.36 \\
            Readability & 65.38 & 0.00 & 3.92 & 5.88 & 90.20 \\
            SimpleReadability & 74.36 & 0.00 & 0.00 & 1.72 & 98.28 \\
            NVAR & 66.67 & 0.00 & 0.00 & 7.69 & 92.31 \\
            NCOMP & 69.23 & 0.00 & 0.00 & 3.70 & 96.30 \\
            Mcclure & 67.95 & 0.00 & 0.00 & 3.77 & 96.23 \\
            McCabe & 69.23 & 0.00 & 0.00 & 3.70 & 96.30 \\
            IndentSTD & 62.82 & 0.00 & 4.08 & 12.24 & 83.67 \\
            MaximumBlockDepth & 69.23 & 0.00 & 0.00 & 5.56 & 94.44 \\
            totalFanOut & 80.77 & 0.00 & 0.00 & 3.17 & 96.83 \\
            Length & 79.49 & 0.00 & 0.00 & 1.61 & 98.39 \\
            MaintainabilityIndex & 80.77 & 0.00 & 0.00 & 1.59 & 98.41 \\
            Parameters & 20.51 & 0.00 & 25.00 & 50.00 & 25.00 \\
            LocalVariables & 65.38 & 0.00 & 1.96 & 7.84 & 90.20 \\
        \bottomrule
    \end{tabular}
    \label{tab:IndividualProjectAnalysisCodeMetrics}

\end{table}

To further understand how \emph{ExtremelyBuggy} methods compare to \emph{Buggy} and \emph{NotBuggy} ones, we also did these comparisons without grouping the \emph{Buggy} and \emph{NotBuggy} methods together. We found that the effect size differences between \emph{NotBuggy} and \emph{Buggy} methods were generally \textit{small} to \textit{medium}. However, the difference between \emph{NotBuggy} and \emph{ExtremelyBuggy} methods was mostly \textit{large} for all code metrics and datasets. This finding confirms the visual findings from figure~\ref{fig:CodeMetricDifs} about the differences between the three bug-proneness types.

\begin{summarybox}{RQ2 Summary}
    At their inception, \emph{ExtremelyBuggy} methods are significantly larger, less readable, and more complex than other methods. This is encouraging because it suggests these methods are distinguishable at the very beginning of their lifetime.
\end{summarybox}

\subsection{RQ3: Predicting the \emph{ExtremelyBuggy} Methods}

In this section, we investigate whether \emph{ExtremelyBuggy} methods can be predicted at the time of their creation. Inspired by prior research~\cite{mo2022exploratory, pascarella_performance_2020, chowdhury_good_2025}, we employ five machine learning models of increasing complexity: Logistic Regression~\cite{hosmer2013applied}, Decision Tree~\cite{song2015decision}, Random Forest~\cite{rigatti2017random}, AdaBoost~\cite{schapire2013explaining}, and a Multi-Layer Perceptron (MLP) model~\cite{murtagh1991multilayer}. Our goal is to systematically evaluate the predictive power of both simple and complex models, assessing whether more sophisticated algorithms offer a significant advantage in identifying \emph{ExtremelyBuggy} methods early in the development lifecycle. All models were implemented using the default parameters provided by the \texttt{scikit-learn} library. This choice was motivated by three factors: (i) Parameter and hyperparameter tuning are computationally expensive under the leave-one-out validation setting—the evaluation approach used in this paper—as it requires training a separate model for each test instance; (ii) We also evaluated the models by randomly changing the parameter and hyperparameter values, and none of which improved the models' accuracy. This also discouraged us from using the more expensive grid search approach. (iii) Prior similar studies found default settings yielded competitive accuracy~\cite{chowdhury_good_2025, pascarella_performance_2020}. Moreover, using default parameters enhances the reproducibility of our study and eases future replication efforts. 

To build and train these models, we first merged the \emph{Buggy} and \emph{NotBuggy} methods into a single category: not \emph{ExtremelyBuggy}. This transformation enables a binary classification task, allowing the models to specifically focus on distinguishing \emph{ExtremelyBuggy} methods from all others. The models were trained using the 14 code metrics computed for each method at the time of its introduction commit. Making predictions by relying solely on metrics available immediately after a method is created allows us to assess the feasibility of detecting \emph{ExtremelyBuggy} methods as early as possible in the software development process.



Due to the highly imbalanced nature of our dataset, we do not use metrics such as \textit{accuracy} to evaluate the performance of the machine learning models, as they could overstate the performance of our models. Instead, we use \textit{precision}, \textit{recall}, \textit{F-measure} (F1), Area Under the Curve (AUC) score, and the Mathews Correlation Coefficient (MCC). AUC ranges from 0 to 1, with a score of 0.5 being no better than random classification and 1 being perfect. MCC ranges from -1 to 1, with 0 being about random classification and a score of 1 being perfect prediction.

We assess prediction performance using a leave-one-out approach at the project level. In each iteration, all methods from one project are used as the test set, while methods from all other projects serve as the training data. This process is repeated for every project, ensuring that each serves as the test set exactly once. This evaluation strategy is motivated by two key factors: (i) it prevents data leakage between training and test sets~\cite{pascarella_performance_2020, chowdhury_method-level_2024}, and (ii) it allows us to assess how well the models generalize across different projects. To address the data imbalance problem—specifically, the significantly smaller number of \emph{ExtremelyBuggy} methods—we apply both undersampling and oversampling techniques using modules provided by the \texttt{scikit-learn} library. While the overall prediction performance remains similar across these strategies, undersampling yields slightly better results.  

\begin{figure}[h]
     \centering
     \begin{subfigure}[b]{0.4\textwidth}
         \centering
         \includegraphics[width=\textwidth]{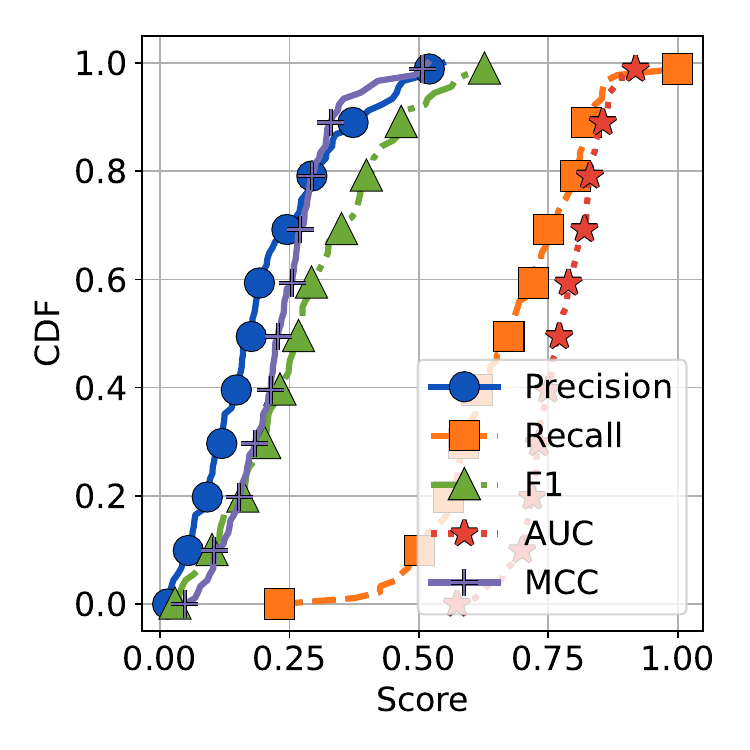}
         \caption{Logistic Regression}
         \label{fig:DecisionTreeRus}
     \end{subfigure}
     \begin{subfigure}[b]{0.4\textwidth}
         \centering
         \includegraphics[width=\textwidth]{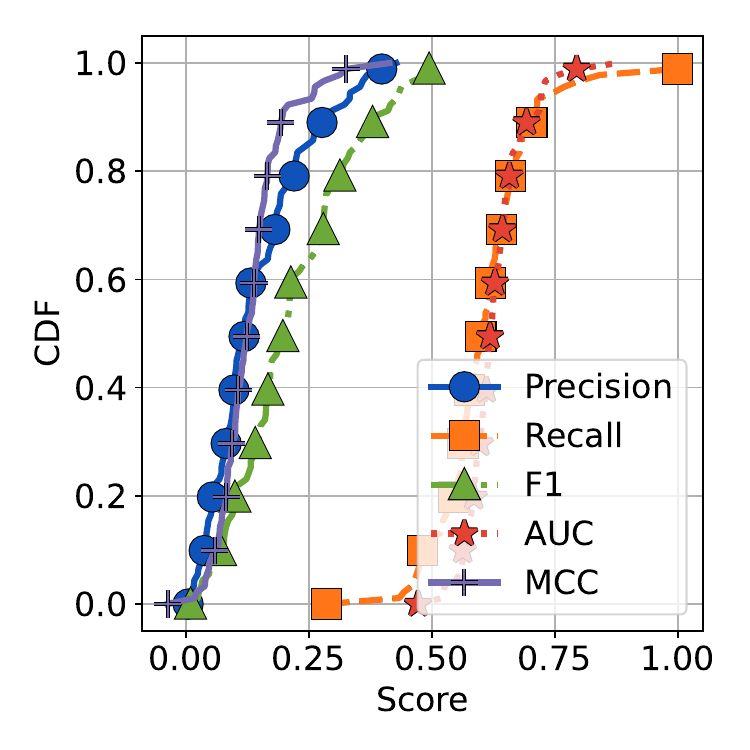}
         \caption{Decision Tree}
         \label{fig:LogRegRus}
     \end{subfigure}
        \caption{Results for two of the select machine learning algorithms for the Balanced dataset. Results are very similar to those of other datasets and algorithms. Results from the other algorithms can be viewed in our publicly shared repository.\protect\footnotemark}
        \Description{Results for two of the select machine learning algorithms for the High Recall dataset.}
        \label{fig:ML_leave_one_out_rus}
\end{figure}

Figure~\ref{fig:ML_leave_one_out_rus} presents various metrics for two selected algorithms using the undersampling approach. The results indicate that recall is substantially higher than precision, suggesting that while the machine learning models are effective at identifying most \emph{ExtremelyBuggy} methods, they tend to generate a very high number of false positives. Most project-level evaluations fall within a similar range, but a few notable outliers stand out. For example, at least one project achieved precision above 0.50, despite approximately 75\% of projects having precision below 0.25. Likewise, although around 90\% of projects achieved recall scores above 0.5, there was at least one project with a recall below 0.25.

\begin{summarybox}{RQ3 Summary}
Predicting \emph{ExtremelyBuggy} methods at inception using simple code metrics and standard machine learning algorithms remains highly challenging. While prediction performance varies across projects, the overall results are not yet reliable enough for practical use.
\end{summarybox}

\footnotetext{\url{https://github.com/SQMLab/ExtremelyBuggyPublicData/tree/main/AdditionalFigures/LeaveOneOutTraining}}
\subsection{RQ4: Observable Characteristics of \emph{ExtremelyBuggy} Methods}


The futility of accurate prediction led us to manually characterize the \emph{ExtremelyBuggy} methods to provide insights for both practitioners and the research community. We employed the thematic analysis~\cite{fereday2006demonstrating} approach to find themes of the \emph{ExtremelyBuggy} methods. Thematic analysis involves identifying general patterns (or "themes") within the data. These themes are created by labeling the data with codes that capture and characterize individual instances of the data and later identifying themes within the codes. Our thematic analysis approach followed the guidelines of Cruzes \textit{et. al}~\cite{CruzesD.S.2011RSfT}, which provides a set of steps for thematic analysis in software engineering research. The approach is described as follows. 

\textbf{Data extraction.} For this analysis, we selected the \emph{ExtremelyBuggy} methods from only the HighPrecision dataset. The reasons behind this are: (i) it contains only 287 unique methods, allowing for thorough manual analysis, and (ii) it has the potential of minimal false positives compared to the other datasets. However, during our review, we still encountered some falsely labeled bugs—mainly due to the inclusion of the keyword \emph{flaw} in commit messages (e.g., “fixing quality flaws”), which sometimes referred to non-functional changes. As a result, we excluded 22 such methods, leaving 265 \emph{ExtremelyBuggy} methods and their evolution history for analysis.

\textbf{Generating initial codes from the methods.} A code is a label that represents a specific, meaningful concept within the data. To minimize internal bias stemming from preconceived assumptions, we employed an open coding approach~\cite{strauss1998basics} to generate codes. Using this approach, we did not begin with predefined codes; instead, we derived codes directly from the data and created new codes whenever the already created codes did not adequately capture the content.


To ensure consistency and correctness in the coding process, the first two authors independently coded 10 selected methods. Following this, the third and fourth authors—who have 2 and 10 years of industry experience, respectively—facilitated a discussion with the first two authors to review the initial codes. We realized that when manually analyzing a data sample, there are multiple dimensions one can take to characterize and assign "codes" to it, as was also discussed in other studies (e.g.,~\cite{jahan2025taxonomy}). This makes it difficult to generate very similar codes by two independent annotators. For example, in our case, one may choose to look at how a method is written, noting it is very large and hard to read, while another may look at the context it is written in, noting the method is used to parse a piece of text. As such, after the discussion session, we agreed to perform our manual analysis in accordance with the following three distinct dimensions:

\begin{enumerate}
    \item \emph{Visual codes.} These codes correspond to how a method looks. When generating these codes, we asked: \textit{What stands out visually when looking at this method (e.g., is it less readable)?}
    \item \emph{Context codes.} These codes represent the context in which a method is used. When generating these codes, we asked: \textit{What is the main purpose of this method (e.g., data handling)?}
    \item \emph{Bug codes.} These codes are generated to represent the reasons for bugs in a method. When generating these codes, we checked the bug fix and asked: \textit{Why did this bug occur (e.g., missing conditionals)?}
\end{enumerate}

The first two authors independently coded an additional set of 10 shared methods and produced highly similar codes. After more verification with the third and fourth authors and growing confidence in labeling consistency, the remaining methods were divided between the first and second authors for individual coding. While it is difficult to directly calculate agreement scores in open coding~\cite{cascio2019team}, as two authors can generate similar codes with different names, the authors conducted a validity check after analyzing all 265 methods. To assess consistency, each author randomly reviewed 30 methods originally coded by the other, and high agreement was observed—differences appeared in only 12 out of 60 cases. Importantly, these were not cases of conflicting codes; rather, one author had identified additional codes that the other had missed. For instance, one author noted the presence of self-admitted technical debt (SATD)~\cite{Potdar:2014, chowdhury_evidence_2025}, which the other did not. As discussed later, all identified codes, including the extra ones, were collaboratively reviewed by the two authors and synthesized into broader themes, ensuring the robustness and comprehensiveness of the analysis.

\textbf{Translating our codes into more general themes.} Once all our methods had been assigned codes, the first two authors collaborated to identify codes with very similar ideas.  They then combined these specific codes into a single, more general theme that accurately encompassed all the codes. It is important to note that a method can belong to multiple themes even within the same theme dimension due to its diverse characteristics.  

\textbf{Creating a model of higher-order themes.} Once the more general themes were created, the two authors then worked together to merge themes into higher-order themes that encompass a variety of different codes and themes. These themes were created with the aim of being generic enough to be present in a significant proportion of the identified methods and accurately quantify what the authors observed in the \emph{ExtremelyBuggy} methods, but not too generic that they lose usefulness in characterizing the \emph{ExtremelyBuggy} methods. Some codes and themes had a very small number of methods that contained them, and did not fit into any of the identified higher-order themes. These codes and themes were placed into a higher-order theme labeled as \emph{Other}. 

\subsubsection{Results}

\begin{figure}[h]
     \centering
     \begin{subfigure}[b]{\textwidth}
         \centering
         \includegraphics[width=0.83\textwidth]{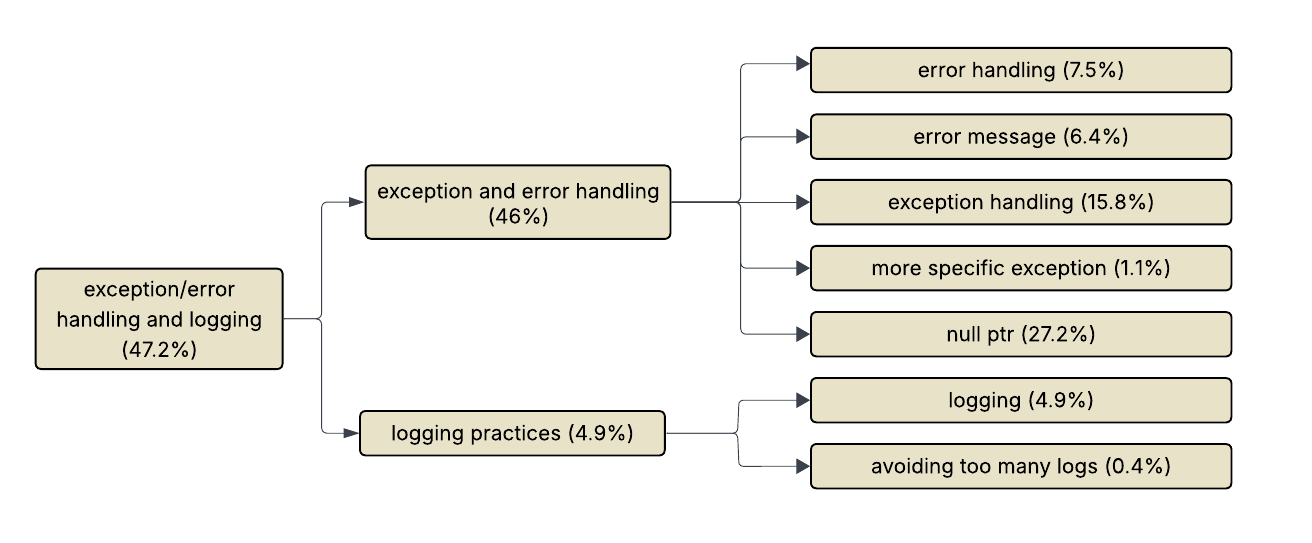}
     \end{subfigure}
        \caption{The taxonomy of one of the themes we identified in the bug themes category during our manual analysis of the \emph{ExtremelyBuggy} methods. The right column shows initial codes, the middle shows more general themes, and the left contains the higher-order theme. Percentages are the number of \emph{ExtremelyBuggy} methods containing that code/theme divided by the total number of \emph{ExtremelyBuggy} methods. The complete taxonomies for all three themes (Bug, Context, and Visual) can be found in the shared repository.\protect\footnotemark}
        \Description{The taxonomy of one of the themes we identified in the bug themes category during our manual analysis}
        \label{fig:ThematicExample}
\end{figure}

In figure~\ref{fig:ThematicExample}, we present an example of the complete thematic analysis process for one of our higher-order themes in the \emph{bug codes} category and the percentages of methods that contain each of the codes and themes. The initial codes that the methods were labeled with are in the right column. In the middle column, we show the initial, more general themes that were identified after looking at the codes. For example, the codes \emph{logging} and \emph{avoiding too many logs} are combined into the more general theme of \emph{logging practices}. Finally, the left column displays the higher-order theme, \emph{exception/error handling and logging}, that was identified using the two themes \emph{exception and error handling} and \emph{logging practices}. We combined logging and exception handling into a single broader category because both are commonly used to detect, respond to, and monitor erroneous behavior in program code. Together, they serve as essential tools for maintaining robustness, facilitating debugging, and ensuring system observability during failure scenarios~\cite{nakshatri2016analysis}. 

\footnotetext{\url{https://github.com/SQMLab/ExtremelyBuggyPublicData/tree/main/AdditionalFigures/ThematicAnalysisTaxonomy}}

For each theme, we calculated the \emph{Count} (the number of methods that include this theme), \emph{Percent} (the percentage of all methods that include this theme), and \emph{Mean Percent} (the average percentage of methods containing this theme across all projects). The \emph{Mean Percent} helps determine whether a high overall percentage is driven by a few outlier projects or represents a more generalizable trend. 


We now describe the themes and show example methods (CodeShovel generated \texttt{JSON} files) associated with the themes. These example files can be viewed in our public repository.~\footnote{\url{https://github.com/SQMLab/ExtremelyBuggyPublicData/tree/main/ManualAnalysisExamples}}

\subsubsection{Visual Themes}

Table~\ref{tab:VisualCodes} shows the final identified themes for the visual/syntactical dimension. These are themes that were identified just by looking at the methods.

\begin{table}[h]
    \centering
    \caption{Final themes from the manual analysis of syntactic and visual characteristics of \textit{ExtremelyBuggy} methods.}
    \begin{tabular}{lccc}
        \toprule
            \textbf{Theme} & \textbf{Count} & \textbf{Percent} & \textbf{Mean Percent} \\
        \midrule
            Confusing Control Flow & 155 & 58.5 & 50.8 \\
            Abnormal Size & 93 & 35.1 & 31.7 \\
            Other & 90 & 34.0 & 40.2 \\
            Poor Readability & 86 & 32.5 & 19.5 \\
            Drastic Change in Size & 80 & 30.2 & 37.6 \\
            SATD & 47 & 17.7 & 13.4 \\
            Sketchy Exception/Error Handling & 35 & 13.2 & 14.8 \\
        \bottomrule
    \end{tabular}
    \label{tab:VisualCodes}
\end{table}

\begin{itemize}
    \item \textbf{Confusing Control Flow.} This theme refers to methods where it is hard to follow the control flow of the method, for example, if the method has many if/else statements or deeply nested code. An example is the processScope() method (105.json) from the intellij-community repository that has many if/else statements and deeply nested code, resulting in complicated bugs related to its control flow.
    \item \textbf{Abnormal Size.} These methods are visually very large. This reinforces the common belief that very large methods are substantially more susceptible to bugs than smaller ones~\cite{gil_correlation_2017,chowdhury_method-level_2024,chowdhury2022empirical}. The processEntry() method (35.json) from the jgit project is an example of a method that is abnormally large.
    \item \textbf{Other.} These are methods that did not show any unique visual characteristics or did not have repeated examples to be grouped into a theme. An example is the disassemble() method (166.json) from the eclipseJdt project. This means that it may be difficult to detect these methods at inception using source code metrics alone.
    \item \textbf{Poor Readability.} Poor Readability refers to methods that are difficult to read or are formatted very poorly. For example, if a method has bad indentation as seen in the search() method (2.json) from the weka project.
    \item \textbf{Drastic Change in Size.} These are methods that started with a few lines of code but got much larger over time. These methods are often unimplemented fully at inception and undergo frequent, large changes that result in bugs being introduced later on. An example is the visitSetOperation() method (18.json) from the presto project.
    \item \textbf{SATD.} Self-admitted technical debt (SATD)~\cite{potdar2014exploratory} occurs when developers acknowledge that there is future work that needs to be done on a method (e.g., with a TODO comment). An example is the parseTransition() method (21.json) from the argouml repository, where unresolved technical debts directly caused bugs in the future.
    \item \textbf{Sketchy Exception/Error Handling.} These methods handled errors and exceptions in dubious ways, often by using many or incomplete try/catch blocks. A notable example is the invoke() method (88.json) that contains many try/catch blocks and try blocks without catch blocks.
\end{itemize}




\subsubsection{Context themes}

Table~\ref{tab:ContextCodes} contains the final 8 themes we identified to explain the contextual purpose of the \emph{ExtremelyBuggy} methods. 

\begin{table}[h]
    \centering
    \caption{Final themes from the manual analysis of the contextual purpose of \textit{ExtremelyBuggy} methods.}
    \begin{tabular}{lccc}
        \toprule
            \textbf{Theme} & \textbf{Count} & \textbf{Percent} & \textbf{Mean Percent} \\
        \midrule
            Core Logic and Algorithms & 70 & 26.4 & 20.6 \\
            Data Handling and Transformation & 56 & 21.1 & 22.1 \\
            Handling External Resources & 44 & 16.6 & 28.8 \\
            Abstract Code Operations & 41 & 15.5 & 9.6 \\
            Error/Exception Handling and Logging & 36 & 13.6 & 9.8 \\
            Initialization and Setup & 27 & 10.2 & 6.3 \\
            UI and Presentation & 24 & 9.1 & 11.0 \\
            Other & 18 & 6.8 & 7.6 \\
        \bottomrule
    \end{tabular}
    \label{tab:ContextCodes}
\end{table}

\begin{itemize}
    \item \textbf{Core Logic and Algorithms.} These methods represent some core logical component of a project that determines how that project works. These methods often have complicated logic with requirements that change as the project changes. For example, the method isInUseableZone() (249.json), from the mage repository performs specific logic to check if an object is in a zone where it can be used. This method gets updated with more and more logic requirements as the project matures which ultimately result in bugs from incorrect logic. 
    \item \textbf{Data Handling and Transformation.} These methods deal with or transform data in some way, for example, by encoding/decoding low-level data or parsing a document. An example is the scanExpr() method (178.json) from the xerces2-j repository, which parses an XPath expression and breaks it up into tokens.
    \item \textbf{Handling External Resources.} External resources such as databases and HTTP messages are handled by these methods. External resources often have specific requirements for how messages are sent and received, which can cause issues if not handled correctly. A notable example is the doPost() method (161.json) from the tomcat project that has bugs resulting from the incorrect formatting of an HTTP POST message.
    \item \textbf{Abstract Code Operations.} These methods perform operations on abstract code representations, such as building and navigating abstract syntax trees (ASTs). However, most of these methods come from a few projects that deal heavily with ASTs. An example is the visit() method (237.json) from the pmd project.
    \item \textbf{Error/Exception Handling and Logging.} The main purpose of these methods is to handle some type of exception/error or to log for debugging purposes. The method handlePackError() (80.json) from the jgit repository is an example of this theme. 
    Note that this theme represents the purpose of the method dealing with error/exception handling, whereas the previous visual theme~\emph{Sketchy Exception/Error Handling} describes how the methods performed exception/error handling.
    \item \textbf{Initialization and Setup.} These methods configure and set up objects, which can often result in developers forgetting some parts that they need to configure or initialize fully. An example is the buildClassifier() method (181.json) from the weka project, where the developers did not account for missing values during configuration, causing bugs.
    \item \textbf{UI and Presentation.} These methods involve the creation and handling of the visual components in a project. One example is the createDropDownList() method (97.json) from the dbeaver repository.
\end{itemize}


\subsubsection{Bug Themes}
In this dimension of themes, we focus on what was done to fix a bug, not what the method is doing (context) or how it looks (visual). Table~\ref{tab:BugCodes} contains the final 10 themes we identified to explain types of bugs in \emph{ExtremelyBuggy} methods. 

\begin{table}[h]
    \centering
    \caption{Final themes from the manual analysis of the reasons for bugs in \textit{ExtremelyBuggy} methods.}
    \begin{tabular}{lccc}
        \toprule
            \textbf{Theme} & \textbf{Count} & \textbf{Percent} & \textbf{Mean Percent} \\
        \midrule
            Conditional Logic & 155 & 58.5 & 54.5 \\
            Exception/Error Handling and Logging & 125 & 47.2 & 43.3 \\
            External Interaction & 73 & 27.5 & 28.8 \\
            Variable Misuse and Naming & 54 & 20.4 & 19.1 \\
            Behavioral Oversights & 43 & 16.2 & 15.2 \\
            Other & 28 & 10.6 & 11.1 \\
            Incorrect Statement Order/Location & 25 & 9.4 & 2.9 \\
            Syntax, Type and Keyword Issues & 22 & 8.3 & 8.6 \\
            Data Lifecycle Mismanagement & 21 & 7.9 & 6.3 \\
            Looping, Indexing and Recursion & 20 & 7.5 & 4.4 \\
        \bottomrule
    \end{tabular}
    \label{tab:BugCodes}
\end{table}

\begin{itemize}
    \item \textbf{Conditional Logic.} These methods had bugs related to incorrect or missing conditional logic. An example is the update() method (109.json) from the mindustry repository.
    \item \textbf{Exception/Error Handling and Logging.} These methods generated incorrect error messages, did not properly handle null pointer exceptions, or used the wrong types of exceptions. An example is getViewSize() (155.json) from the wicket repository. 
    \item \textbf{External Interaction.} These types of bugs occur when a method interacts with other methods and resources. These bugs commonly occur from calling the wrong method, often due to it being overloaded or having a similar name to the one it is trying to call. An example is the method addDefaultAttributes() (296.json) from the xerces2-j project that calls toString() instead of stringValue(). 
    \item \textbf{Variable Misuse and Naming.} For these methods, bug-fixes were related to the incorrect use and naming of variables. The method lookupValues() (44.json) from the pentaho-kettle repository provides an example where the wrong variable was passed to a method.
    \item \textbf{Behavioral Oversights.} This theme represents bugs that were caused by some missed edge case or behavior that was not handled properly. For example, an edge case was missed in the getHadoopUser() method (7.json) from the flink repository.
    \item \textbf{Incorrect Statement Order/Location.} This type of bug occurs when the correct statements are present, but are in the wrong order or location in the method. This can often occur due to deep nesting, resulting in a statement being in the wrong set of brackets. The method readHeaders() (145.json) from the netty project is one example.
    \item \textbf{Syntax, Type and Keyword Issues.} These bugs occur due to typing issues, usually only requiring one keyword or type to be changed to be fixed. We also noticed that variables were often cast to the incorrect type, which caused the wrong methods to be called or were unable to be called. An example is the method supresses() (242.json) from the pmd repository.
    \item \textbf{Data Lifecycle Mismanagement.} These are bugs related to the creation, destruction, and handling of data. One example is in the failed() (59.json) method from the pulsar repository that missed releasing resources, causing a memory leak.
    \item \textbf{Looping, Indexing and Recursion.} These are bugs related to any kind of looping or recursion and indexing problems. For example, in the replaceEvent() method (328.json) from the mage repo, there is a missing break statement in a loop.
\end{itemize}

\begin{summarybox}{RQ4 Summary}
 \emph{ExtremelyBuggy} methods often exhibit distinct visual characteristics (e.g., code smells such as large size or poor readability) and have frequent contextual roles (e.g., handling core logic). The underlying causes of bugs—such as missing edge cases or incorrect exception handling—are also commonly shared among these methods. These observations can help practitioners identify areas that warrant greater attention and support researchers in designing more effective features for future machine learning–based bug prediction models.
\end{summarybox}
\section{Discussion}

Due to the importance of method-level bug prediction, significant research has focused on this topic in recent years~\cite{mo2022exploratory, giger_method-level_2012,chowdhury_method-level_2024, Shippey, ferenc_automatically_2020, pascarella_performance_2020}. However, these earlier studies treated all bug-prone methods as the same, regardless of how many times a method was involved in bug fixes. In this study, we choose to explore the \emph{ExtremelyBuggy} methods as understanding and capturing these methods would enable more optimized resource allocation. We found that \emph{ExtremelyBuggy} methods are very small in numbers but can often account for a large number of bugs in a project (\textbf{RQ1}). This is encouraging because early optimization of these small numbers of methods can significantly reduce future maintenance burdens. Our analysis showed that \emph{ExtremelyBuggy} methods are significantly larger, more complex, and less readable than both \emph{Buggy} and \emph{NotBuggy} methods (\textbf{RQ2}). Despite these differences, \emph{ExtremelyBuggy} methods are very challenging to predict at their inception, as machine learning models showed poor prediction performance (\textbf{RQ3}). However, with manual analysis, we found that the \emph{ExtremelyBuggy} methods have common visual characteristics, contextual purposes, and similar root causes of bugs (\textbf{RQ4}). 

\subsection{Implications for Researchers}
For the research community, we provide a dataset of over 1.25 million Java methods originating from 98 popular open-source projects. To the best of our knowledge, this is the largest dataset on method-level bugs. This enormous dataset not only enables replicating our results, but also can help answer more research questions. While manually analyzing the \emph{ExtremelyBuggy} methods (\textbf{RQ4}), we discovered that over 45\% of the bugs in the \emph{ExtremelyBuggy} methods were introduced after modification to the original code. This not only explains the poor prediction performance at the inception of the \emph{ExtremelyBuggy} methods, but also suggests that future research should incorporate the change history of methods while building the prediction models.

Many of the common themes we observed are not detectable from code metrics alone---e.g., HTTP and database operations, SATD, and abstract code operations. This implies that incorporating code embeddings as additional features might help improve the accuracy of these models---this, however, will be an expensive process given the enormous amount of methods we have. We also noticed that prediction performance is not uniform across all projects, as some projects exhibited unusually high or low accuracy. This implies that future research can focus on optimal project set selection for training models based on the project under test. Previous work in similar studies has shown the promise of optimal project selection during model training~\cite{chowdhury_method-level_2024, sun2021cfps}.

One of the major challenges in method-level bug prediction~\cite{chowdhury_method-level_2024}, and thus in predicting \emph{ExtremelyBuggy} methods, is the presence of tangled commits. Although a significant amount of research~\cite{muylaert_research_2018, partachi_flexeme_2020,li_utango_2022,xu_detecting_2025} has focused on untangling tangled code changes, none of these approaches addresses the problem at the method-level granularity. Only recently, Opu et al.~\cite{opu2025llm} evaluated the effectiveness of Large Language Models for detecting tangling in method-level bug-fix commits and demonstrated significant success. Future research should investigate whether applying untangling as a preprocessing step can further improve the accuracy reported in our study.

Another challenge we encountered was the data imbalance issue, as there were very few \emph{ExtremelyBuggy} samples compared to other classes. This is a well-known problem that makes robust model training particularly challenging~\cite{Manahel:2025}. Although we attempted to mitigate this issue using simple oversampling and undersampling techniques, future research could explore more advanced RAG-based augmentation approaches~\cite{daneshvar2025vulscriber} to improve minority-class representation and overall model robustness.

\subsection{Implications for Practitioners}


Practitioners seeking optimization opportunities should take into account the contextual and visual themes identified in this paper. For instance, methods associated with a project's \emph{core logic and algorithms} are significantly more likely to become \emph{ExtremelyBuggy}. When designing solutions for such methods, conducting a focused brainstorming session can help identify the most effective implementation strategies and anticipate potential future requirement changes related to the core logic that might introduce bugs.

Notably, \emph{conditional logic} emerged as one of the most bug-prone elements in methods. Therefore, when developers encounter complex conditional structures—especially in high-risk contexts—they should consider breaking them down into simpler, smaller logical components. Additionally, practitioners are encouraged to use code smell detectors, as many extremely buggy methods exhibit common code smells. Practitioners should also address self-admitted technical debts (SATD) as early as possible. In addition, dedicated effort to the design and implementation of test code can help prevent bugs related to unhandled edge cases and incorrect exception handling that we observed frequently with these methods.

\section{Threats to Validity}

We have identified multiple threats that may impact the validity of our findings.

\emph{Construct validity} is impacted by our choice of the CodeShovel tool, which is not 100\% accurate while extracting method histories. 
We used a variety of popular product metrics for evaluating the \emph{ExtremelyBuggy} methods; however, other metrics could be considered. For example, developer-centric metrics could be used. In this paper, we chose not to include developer-centric metrics as they are not very useful at the beginning of a project, and we wanted our findings to be applicable at a project's inception. To detect the bug-proneness of a method, we used a keyword-based approach, which is known to suffer from false positives~\cite{bird2009fair}. However, this threat was mitigated, to some extent, by the construction of three different datasets: one with high precision, one with high recall, and one aimed at a balance between precision and recall.

For the manual analysis of the methods, the HighPrecision dataset was used to reduce the number of false positives during labeling. However, the HighPrecision dataset can only consider a commit to be a bug-fix if there were no tangled changes, that is, if only one method was changed in that commit. This may introduce bias towards certain types of bugs that only affect one method.



\emph{Internal validity} is hampered due to our selection of statistical tests such as Kendall's $\tau$, the Wilcoxon Rank Sum test, and Cliff's \textit{d}. However, these tests are well established and commonly used in software engineering research \cite{chowdhury_revisiting_2022, gil_correlation_2017, inozemtseva2014coverage}. To identify themes in the \emph{ExtremelyBuggy} methods, we used a manual labeling approach, which is inherently susceptible to human bias and error. To mitigate this threat, we compared the labeling of the first two authors and also received help from the third and fourth authors with 2+ and 10+ years of industry experience, respectively.

\emph{External validity} is affected by our use of only open-source Java repositories, and it is unclear how our findings would extend to closed-source projects or projects in other languages. Also, for our analysis, we had to ensure that the descriptions (e.g., commit messages) for our repositories were written in English, and discarded other repositories with non-English descriptions. 


\emph{Conclusion validity} of this study is influenced by all previous threats to validity. Additionally, our selection of a five-year threshold for age-normalization can impact the conclusion we made. To mitigate this threat, we repeated most of our experiments using a two and eight-year threshold, and none of the major results from the first three RQs changed. However, future work is needed to see how the results may change for the manual analysis of \emph{ExtremelyBuggy} methods using different age-normalization thresholds.

\section{Conclusion}
This study provides the first large-scale, method-level investigation into \textit{ExtremelyBuggy} methods using a dataset of over 1.25 million methods from 98 open-source Java systems. Our empirical analysis demonstrates that although \textit{ExtremelyBuggy} methods constitute only a small fraction of a project’s methods, they frequently account for a disproportionately large share of bugs. These methods exhibit clear distinguishing characteristics at their inception: they are typically larger, more complex, less readable, and less maintainable than both simple buggy and non-buggy methods. Yet, despite these measurable differences, our evaluation of multiple machine learning models reveals that early prediction of \textit{ExtremelyBuggy} methods remains highly challenging due to tangled code changes, severe class imbalance, cross-project variability, and the fact that many defects arise through later code evolution rather than initial implementation. To complement the quantitative analyses, our manual thematic examination of 265 \textit{ExtremelyBuggy} methods reveals recurring visual, contextual, and defect-related patterns. Visually, these methods often exhibit clear code smells, including confusing control flow, abnormal size growth, poor readability, self-admitted technical debt, and fragile exception-handling structures.

These findings emphasize the need for richer representations of code and project history and the necessity of more research on how to deal with class imbalance and tangled changes. Overall, our work deepens the understanding of extreme bug-prone methods and establishes a foundation for future research toward more robust, early identification of methods likely to cause recurring bugs. 


\begin{acks}
This research project was supported by an NSERC Discovery Grant, Canada. 
\end{acks}

\printbibliography

\appendix

\end{document}